\documentclass{article}
\usepackage{amssymb}

\input{tcilatex}

\begin{document}

\author{Sawa Manoff \\
\textit{Institute for Nuclear Research and Nuclear Energy}\\
\textit{\ Department of Theoretical Physics}\\
\textit{\ Blvd. Tzarigradsko Chaussee 72}\\
\textit{\ 1784 Sofia - Bulgaria}}
\date{\textit{E-mail address: smanov@inrne.bas.bg }}
\title{\textbf{Centrifugal (centripetal), Coriolis' velocities,
accelerations and Hubble's law \ in spaces with affine connections and
metrics}}
\maketitle

\begin{abstract}
\textit{The notions of centrifugal (centripetal) and Coriolis' velocities
and accelerations are introduced and considered in spaces with affine
connections and metrics [}$(\overline{L}_{n},g)$-\textit{spaces] as
velocities and accelerations of flows of mass elements (particles) moving in
space-time. It is shown that these types of velocities and accelerations are
generated by the relative motions between the mass elements. They are
closely related to the kinematic characteristics of the relative velocity
and relative acceleration. The null (isotropic) vector fields are considered
and their relations with the centrifugal (centripetal) velocity are
established. The centrifugal (centripetal) velocity is found to be in
connection with the Hubble law and the generalized Doppler effect in spaces
with affine connections and metrics. The centrifugal (centripetal)
acceleration could be interpreted as gravitational acceleration as it has
been done in the Einstein theory of gravitation. This fact could be used as
a basis for working out of new gravitational theories in spaces with affine
connections and metrics. }
\end{abstract}

\section{Introduction}

1. The relative velocity and the relative acceleration between particles or
mass elements of a flow are important characteristics for describing the
evolution and the motion of a dynamic system. On the other side, the
kinematic characteristics related to the relative velocity and the relative
acceleration (such as deformation velocity and acceleration, shear velocity
and acceleration, rotation velocity and acceleration, and expansion velocity
and acceleration) characterize specific relative motions of particles and
/or mass elements in a flow \cite{Manoff-0}$\div $\cite{Manoff-4}. On the
basis of the links between the kinematic characteristics related to the
relative velocity and these related to the relative acceleration the
evolution of a system of particles or mass elements of a flow could be
connected to the geometric properties of the corresponding mathematical
model of a space or space-time \cite{Stephani}. Many of the notions of
classical mechanics or of classical mechanics of continuous media preserve
their physical interpretation in more comprehensive spaces than the
Euclidean or Minkowskian spaces, considered as mathematical models of space
or space-time \cite{Ehlers}. On this background, the generalizations of the
notions of Coriolis' and centrifugal (centripetal) accelerations \cite%
{Lammerzahl} from classical mechanics in Euclidean spaces are worth to be
investigated in spaces with affine connections and metrics [$(\overline{L}%
_n,g)$-spaces] \cite{Bishop} $\div $ \cite{Manoff-7a}.

2. Usually, the Coriolis' and centrifugal (centripetal) accelerations are
considered as apparent accelerations, generated by the non-inertial motion
of the basic vector fields determining a co-ordinate system or a frame of
reference in an Euclidean space. In Einstein's theory of gravitation (ETG)
these types of accelerations are considered to be generated by a symmetric
affine connection (Riemannian connection) compatible with the corresponding
Riemannian metric \cite{Stephani}. In both cases they are considered as
corollaries of the non-inertial motion of particles (i.e. of the motion of
particles in non-inertial co-ordinate system).

3. In the present paper the notions of Coriolis' and centrifugal
(centripetal) accelerations are considered with respect to their relations
with the geometric characteristics of the corresponding models of space or
space-time. It appears that accelerations of these types are closely related
to the kinematic characteristics of the relative velocity and of the
relative acceleration. The main idea is to be found \ out how a Coriolis' or
centrifugal (centripetal) acceleration acts on a mass element of a flow or
on a single particle during its motion in space or space-time described by a
space of affine connections and metrics. In Section 1 the notions of
centrifugal (centripetal) and Coriolis' velocities are introduced and
considered in $(\overline{L}_n,g)$-spaces. The relations between the
kinematic characteristics of the relative velocity and the introduced
notions are established. The null (isotropic) vector fields are considered
and their relations with the centrifugal (centripetal) velocity are
established. The centrifugal (centripetal) velocity is found to be in
connection with the Hubble law and the generalized Doppler effect in spaces
with affine connections and metrics. In Section 2 the notions of centrifugal
(centripetal) and Coriolis' accelerations are introduced and their relations
to the kinematic characteristics of the relative acceleration are found. In
Section 3 the interpretation of the centrifugal (centripetal) acceleration
as gravitational acceleration is given and illustrated on the basis of the
Einstein theory of gravitation and especially on the basis of the
Schwarzschild metric in vacuum.

4. The main results in the paper are given in details (even in full details)
for these readers who are not familiar with the considered problems. The
definitions and abbreviations are identical to those used in \cite{Manoff-5}%
, \cite{Manoff-7}, \cite{Manoff-7a}. The reader is kindly asked to refer to
them for more details and explanations of the statements and results only
cited in this paper.

\section{Centrifugal (centripetal) and Coriolis' velocities}

Let us now recall some well known facts from kinematics of vector fields
over spaces with affine connections and metrics [$(\overline{L}_n,g)$ -
spaces], considered as models of space or space-time \cite{Manoff-0}$\div $%
\cite{Manoff-7a}.

Every vector field $\xi $ could be represented by the use of a non-isotropic
(non-null) vector field $u$ and its corresponding contravariant and
covariant projective metrics $h^{u}$ and $h_{u}$ in the form 
\begin{equation}
\xi =\frac{1}{e}\cdot g(u,\xi )\cdot u+\overline{g}[h_{u}(\xi )]\text{ ,}
\label{1.1}
\end{equation}
where

\begin{eqnarray}
h^{u} &=&\overline{g}-\frac{1}{e}\cdot u\otimes u\text{ ,~\ \ \ \ \ \ }%
h_{u}=g-\frac{1}{e}\cdot g(u)\otimes g(u)\text{ ,}  \label{1.2} \\
\overline{g} &=&g^{ij}\cdot e_{i}.e_{j}\text{ , \ \ \ }g^{ij}=g^{ji}\text{ ,
\ \ }e_{i}.e_{j}=\frac{1}{2}\cdot (e_{i}\otimes e_{j}+e_{j}\otimes e_{i})%
\text{ ,}  \nonumber \\
g &=&g_{ij}\cdot e^{i}.e^{j}\text{ , \ \ \ \ }g_{ij}=g_{ji}\text{ \ , \ \ }%
e^{i}.e^{j}=\frac{1}{2}\cdot (e^{i}\otimes e^{j}+e^{j}\otimes e^{i})\text{
,\ }  \nonumber
\end{eqnarray}

\begin{eqnarray}
e &=&g(u,u)=g_{\overline{k}\overline{l}}\cdot u^{k}\cdot u^{l}=u_{\overline{k%
}}\cdot u^{k}=u_{k}\cdot u^{\overline{k}}:\neq 0\text{ , \ }  \label{1.2a} \\
g(u,\xi ) &=&g_{\overline{i}\overline{j}}\cdot u^{i}\cdot \xi ^{j}\text{ , }%
e_{i}=\partial _{i}\text{ \ , \ \ \ }e^{j}=dx^{j}\text{ \ in a co-ordinate
basis.}  \label{1.2b}
\end{eqnarray}

By means of the representation of $\xi $ the notions of relative velocity
and relative acceleration are introduced \cite{Manoff-5}, \cite{Manoff-7} in 
$(\overline{L}_n,g)$ - spaces. By that, the vector field $\xi $ has been
considered as vector field orthogonal to the vector field $u$, i.e. $g(u,\xi
)=0$, $\xi =\xi _{\perp }=\overline{g}[h_u(\xi )]$. Both the fields are
considered as tangent vector fields to the corresponding co-ordinates, i.e.
they fulfil the condition \cite{Bishop} $\pounds _\xi u=-\pounds _u\xi
=[u,\xi ]=0$, where $[u,\xi ]=u\circ \xi -\xi \circ u$. Under these
preliminary conditions the relative velocity $_{rel}v$ could be found in the
form \cite{Manoff-5}, \cite{Manoff-7} 
\begin{equation}
_{rel}v=\overline{g}[d(\xi _{\perp })]\text{ \ ,}  \label{1.3}
\end{equation}
where 
\begin{equation}
d=\sigma +\omega +\frac 1{n-1}\cdot \theta \cdot h_u\text{ \ .}  \label{1.4}
\end{equation}

The tensor $d$ is the deformation velocity tensor; the tensor $\sigma $ is
the shear velocity tensor; the tensor $\omega $ is the rotation velocity
tensor; the invariant $\theta $ is the expansion velocity invariant. The
vector field $_{rel}v$ is interpreted as the relative velocity of two points
(mass elements, particles) moving in a space or space-time and having equal
proper times \cite{Manoff-0}$\div $\cite{Manoff-4}. The vector field $\xi
_{\perp }$ is orthogonal to $u$ and is interpreted as deviation vector
connecting the two mass elements (particles) (if considered as an
infinitesimal vector field).

Let us now consider the representation of the relative velocity by the use
of a non-isotropic (non-null) vector field $\xi _{\perp }$ (orthogonal to $u$%
) and its corresponding projective metrics 
\begin{eqnarray}
h^{\xi _{\perp }} &=&\overline{g}-\frac{1}{g(\xi _{\perp },\xi _{\perp })}%
\cdot \xi _{\perp }\otimes \xi _{\perp }\text{ \ ,}  \label{1.5} \\
h_{\xi _{\perp }} &=&g-\frac{1}{g(\xi _{\perp },\xi _{\perp })}\cdot g(\xi
_{\perp })\otimes g(\xi _{\perp })\text{ \ .}  \label{1.6}
\end{eqnarray}

Then a vector field $v$ could be represented in the form 
\begin{equation}
v=\frac{g(v,\xi _{\perp })}{g(\xi _{\perp },\xi _{\perp })}\cdot \xi _{\perp
}+\overline{g}[h_{\xi _{\perp }}(v)]\text{ \ .}  \label{1.7}
\end{equation}

Therefore, the relative velocity $_{rel}v$ could be now written in the form 
\begin{equation}
_{rel}v=\frac{g(_{rel}v,\xi _{\perp })}{g(\xi _{\perp },\xi _{\perp })}\cdot
\xi _{\perp }+\overline{g}[h_{\xi _{\perp }}(_{rel}v)]=v_{z}+v_{c}\text{ \ \
,}  \label{1.8}
\end{equation}
where 
\begin{equation}
v_{z}=\frac{g(_{rel}v,\xi _{\perp })}{g(\xi _{\perp },\xi _{\perp })}\cdot
\xi _{\perp }\text{ \ , \ \ \ \ \ \ \ }v_{c}=\overline{g}[h_{\xi _{\perp
}}(_{rel}v)]\text{ \ .}  \label{1.9}
\end{equation}

The vector field $v_{z}$ is collinear to the vector field $\xi _{\perp }$.
If the factor (invariant) before $\xi _{\perp }$ is positive, i.e. if 
\begin{equation}
\frac{g(_{rel}v,\xi _{\perp })}{g(\xi _{\perp },\xi _{\perp })}>0\text{ }
\label{1.10}
\end{equation}
the vector field $v_{z}$ is called \textit{centrifugal velocity}. If the
factor (invariant) before $\xi _{\perp }$ is negative, i.e. if 
\begin{equation}
\frac{g(_{rel}v,\xi _{\perp })}{g(\xi _{\perp },\xi _{\perp })}<0
\label{1.11}
\end{equation}
the vector field $v_{z}$ is called \textit{centripetal velocity}.

The vector field $v_{c}$ is called \textit{Coriolis' velocity.}

\subsection{Centrifugal (centripetal) velocity}

\paragraph{Properties of the centrifugal (centripetal) velocity}

(a) Since $v_{z}$ is collinear to $\xi _{\perp }$, it is orthogonal to the
vector field $u$, i.e. 
\begin{equation}
g(u,v_{z})=0\text{ \ .}  \label{1.12}
\end{equation}

(b) The centrifugal (centripetal) velocity $v_{z}$ is orthogonal to the
Coriolis velocity $v_{c}$%
\begin{equation}
g(v_{z},v_{c})=0\text{ .}  \label{1.13}
\end{equation}

(c) The length of the vector $v_{z}$ could be found by means of the relation 
\begin{eqnarray}
g(v_{z},v_{z}) &=&v_{z}^{2}=g(\frac{g(_{rel}v,\xi _{\perp })}{g(\xi _{\perp
},\xi _{\perp })}\cdot \xi _{\perp },\frac{g(_{rel}v,\xi _{\perp })}{g(\xi
_{\perp },\xi _{\perp })}\cdot \xi _{\perp })=  \label{1.14} \\
&=&\frac{[g(_{rel}v,\xi _{\perp })]^{2}}{g(\xi _{\perp },\xi _{\perp })^{2}}%
\cdot g(\xi _{\perp },\xi _{\perp })=\frac{[g(_{rel}v,\xi _{\perp })]^{2}}{%
g(\xi _{\perp },\xi _{\perp })}=\frac{[g(_{rel}v,\xi _{\perp })]^{2}}{\xi
_{\perp }^{2}}\text{ \ .}  \nonumber
\end{eqnarray}

From the last expression we can conclude that the square of the length of
the centrifugal (centripetal) velocity is in general in inverse proportion
to the length of the vector field $\xi _{\perp }$.

\textit{Special case:} $M_n:=E_n$, $n=3$ (3-dimensional Euclidean space): $%
\xi _{\perp }=\overrightarrow{r}$, $g(\xi _{\perp },\xi _{\perp })=r^2$%
\begin{equation}
v_z^2=\frac{[g(\overrightarrow{r},_{rel}\overrightarrow{v})]^2}{r^2}\text{ ,
\thinspace \thinspace \thinspace \thinspace \thinspace \thinspace \thinspace
\thinspace \thinspace \thinspace \thinspace \thinspace \thinspace \thinspace
\thinspace \thinspace }l_{v_z}=\sqrt{v_z^2}=\frac{g(\overrightarrow{r},_{rel}%
\overrightarrow{v})}r\text{ \thinspace \thinspace \thinspace \thinspace .}
\label{1.15}
\end{equation}

If the relative velocity $_{rel}v$ is equal to zero then $v_{z}=0$.

(d) The scalar product between $v_{z}$ and $_{rel}v$ could be found in its
explicit form by the use of the explicit form of the relative velocity $%
_{rel}v$%
\begin{equation}
_{rel}v=\overline{g}[d(\xi _{\perp })]=\overline{g}[\sigma (\xi _{\perp })]+%
\overline{g}[\omega (\xi _{\perp })]+\frac{1}{n-1}\cdot \theta \cdot \xi
_{\perp }\text{ ,}  \label{1.16}
\end{equation}
and the relations 
\begin{equation}
g(\xi _{\perp },_{rel}v)=g(\xi _{\perp },\overline{g}[\sigma (\xi _{\perp
})])+\overline{g}(\xi _{\perp },\overline{g}[\omega (\xi _{\perp })])+\frac{1%
}{n-1}\cdot \theta \cdot g(\xi _{\perp },\xi _{\perp })\text{ ,}
\label{1.17}
\end{equation}
\begin{eqnarray}
g(\xi _{\perp },\overline{g}[\sigma (\xi _{\perp })]) &=&\sigma (\xi _{\perp
},\xi _{\perp })\text{ ,}  \label{1.18} \\
\overline{g}(\xi _{\perp },\overline{g}[\omega (\xi _{\perp })]) &=&\omega
(\xi _{\perp },\xi _{\perp })=0\text{ ,}  \label{1.19} \\
g(\xi _{\perp },_{rel}v) &=&\sigma (\xi _{\perp },\xi _{\perp })+\frac{1}{n-1%
}\cdot \theta \cdot g(\xi _{\perp },\xi _{\perp })\text{ .}  \label{1.20}
\end{eqnarray}

For 
\begin{eqnarray}
g(v_{z},_{rel}v) &=&g(\frac{g(\xi _{\perp },_{rel}v)}{g(\xi _{\perp },\xi
_{\perp })}.\xi _{\perp },_{rel}v)=\frac{g(\xi _{\perp },_{rel}v)}{g(\xi
_{\perp },\xi _{\perp })}\cdot g(\xi _{\perp },_{rel}v)=  \nonumber \\
&=&\frac{[g(\xi _{\perp },_{rel}v)]^{2}}{g(\xi _{\perp },\xi _{\perp })}%
=v_{z}^{2}\text{ ,}  \label{1.21}
\end{eqnarray}
it follows that 
\begin{equation}
g(v_{z},_{rel}v)=g(v_{z},v_{z})=v_{z}^{2}\text{ .}  \label{1.21a}
\end{equation}

\textit{Remark.} $g(v_{z},_{rel}v)=g(v_{z},v_{z})=v_{z}^{2}$ because of $%
g(v_{z},_{rel}v)=g(v_{z},v_{z}+v_{c})=g(v_{z},v_{z})+g(v_{z},v_{c})=g(v_{z},v_{z}) 
$.

On the other side, we can express $v_{z}^{2}$ by the use of the kinematic
characteristics of the relative velocity $\sigma $, $\omega $, and $\theta $%
\begin{eqnarray}
v_{z}^{2} &=&\frac{[g(\xi _{\perp },_{rel}v)]^{2}}{g(\xi _{\perp },\xi
_{\perp })}=\frac{1}{g(\xi _{\perp },\xi _{\perp })}\cdot \lbrack \sigma
(\xi _{\perp },\xi _{\perp })+\frac{1}{n-1}\cdot \theta \cdot g(\xi _{\perp
},\xi _{\perp })]^{2}=  \nonumber \\
&=&\frac{1}{g(\xi _{\perp },\xi _{\perp })}\cdot \{[\sigma (\xi _{\perp
},\xi _{\perp })]^{2}+\frac{1}{(n-1)^{2}}\cdot \theta ^{2}\cdot \lbrack
g(\xi _{\perp },\xi _{\perp })]^{2}+  \nonumber \\
&&+\frac{2}{n-1}\cdot \theta \cdot \sigma (\xi _{\perp },\xi _{\perp })\cdot
g(\xi _{\perp },\xi _{\perp })\}\text{ ,}  \nonumber \\
v_{z}^{2} &=&\frac{[\sigma (\xi _{\perp },\xi _{\perp })]^{2}}{g(\xi _{\perp
},\xi _{\perp })}+\frac{1}{(n-1)^{2}}\cdot \theta ^{2}\cdot g(\xi _{\perp
},\xi _{\perp })+\frac{2}{n-1}\cdot \theta \cdot \sigma (\xi _{\perp },\xi
_{\perp })\text{ .}  \label{1.22}
\end{eqnarray}

\textit{Special case:} $\theta :=0$ (expansion-free relative velocity) 
\begin{equation}
v_z^2=\frac{[\sigma (\xi _{\perp },\xi _{\perp })]^2}{g(\xi _{\perp },\xi
_{\perp })}=\frac{[\sigma (\xi _{\perp },\xi _{\perp })]^2}{\xi _{\perp }^2}=%
\frac{[\sigma (\xi _{\perp },\xi _{\perp })]^2}{\mp l_{\xi _{\perp }}^2}%
\text{ ,}  \label{1.23}
\end{equation}
\begin{eqnarray}
g(\xi _{\perp },\xi _{\perp }) &=&\mp l_{\xi _{\perp }}^2\text{ \thinspace
\thinspace ,\thinspace \thinspace \thinspace \thinspace }  \label{1.23a} \\
g(\xi _{\perp },\xi _{\perp }) &=&-l_{\xi _{\perp }}^2\text{\thinspace
\thinspace }\,\,\,\text{if \thinspace \thinspace \thinspace \thinspace
\thinspace \thinspace }g(u,u)=e=+l_u^2\text{ \thinspace \thinspace ,}
\label{1.23b} \\
g(\xi _{\perp },\xi _{\perp }) &=&+l_{\xi _{\perp }}^2\text{\thinspace }%
\,\,\,\,\,\text{if\thinspace \thinspace \thinspace \thinspace \thinspace
\thinspace \thinspace \thinspace }g(u,u)=e=-l_u^2\text{ \thinspace
\thinspace \thinspace .\thinspace \thinspace \thinspace \thinspace }
\label{1.23c}
\end{eqnarray}

\textit{Special case: }$\sigma :=0$ (shear-free relative velocity) 
\begin{equation}
v_z^2=\frac 1{(n-1)^2}\cdot \theta ^2\cdot g(\xi _{\perp },\xi _{\perp })%
\text{ .}  \label{1.23d}
\end{equation}

(e) The explicit form of $v_{z}$ could be found as 
\begin{eqnarray}
v_{z} &=&\frac{g(_{rel}v,\xi _{\perp })}{g(\xi _{\perp },\xi _{\perp })}%
\cdot \xi _{\perp }=\frac{1}{g(\xi _{\perp },\xi _{\perp })}\cdot \lbrack
\sigma (\xi _{\perp },\xi _{\perp })+\frac{1}{n-1}\cdot \theta \cdot g(\xi
_{\perp },\xi _{\perp })]\cdot \xi _{\perp }=  \nonumber \\
&=&[\frac{\sigma (\xi _{\perp },\xi _{\perp })}{g(\xi _{\perp },\xi _{\perp
})}+\frac{1}{n-1}\cdot \theta ]\cdot \xi _{\perp }\text{ ,}  \nonumber \\
v_{z} &=&[\frac{1}{n-1}\cdot \theta +\frac{\sigma (\xi _{\perp },\xi _{\perp
})}{g(\xi _{\perp },\xi _{\perp })}]\cdot \xi _{\perp }\text{ .}
\label{1.24}
\end{eqnarray}

If 
\begin{equation}
\frac{1}{n-1}\cdot \theta +\frac{\sigma (\xi _{\perp },\xi _{\perp })}{g(\xi
_{\perp },\xi _{\perp })}>0\text{ , \ \ \ \ \ \ \ \ }\xi _{\perp }^{2}=g(\xi
_{\perp },\xi _{\perp })\text{ ,}  \label{1.25}
\end{equation}
we have a centrifugal (relative) velocity.

If 
\begin{equation}
\frac 1{n-1}\cdot \theta +\frac{\sigma (\xi _{\perp },\xi _{\perp })}{g(\xi
_{\perp },\xi _{\perp })}<0\text{ ,}  \label{1.26}
\end{equation}
we have a centripetal (relative) velocity.

\textit{Special case: }$\theta :=0$ (expansion-free relative velocity) 
\begin{equation}
v_z=\frac{\sigma (\xi _{\perp },\xi _{\perp })}{g(\xi _{\perp },\xi _{\perp
})}\cdot \xi _{\perp }\text{ .}  \label{1.26a}
\end{equation}

\textit{Special case:} $\sigma :=0$ (shear-free relative velocity) 
\begin{equation}
v_{z}=\frac{1}{n-1}\cdot \theta \cdot \xi _{\perp }\text{ \ .}  \label{1.27}
\end{equation}

If the expansion invariant $\theta >0$ we have centrifugal (or expansion)
(relative) velocity. If the expansion invariant $\theta <0$ we have
centripetal (or contraction) (relative) velocity. Therefore, in the case of
a shear-free relative velocity $v_{z}$ is proportional to the expansion
velocity invariant $\theta $.

If we introduce the vector field $n_{\perp }$, normal to $u$ and normalized,
as 
\begin{eqnarray}
n_{\perp } &:&=\frac{\xi _{\perp }}{l_{\xi _{\perp }}}\text{ \ , \thinspace
\ \ \ \ }g(\xi _{\perp },\xi _{\perp })=\mp l_{\xi _{\perp }}^2\text{ \ ,
\thinspace \thinspace \thinspace \thinspace }\xi _{\perp }=l_{\xi _{\perp
}}\cdot n_{\perp }\text{ \thinspace \thinspace \thinspace \thinspace ,\ \ \
\ \ \ \ }  \label{1.28a} \\
l_{\xi _{\perp }} &=&\,\mid g(\xi _{\perp },\xi _{\perp })\mid ^{1/2}\text{
, \ \ \ }g(n_{\perp },n_{\perp })=\mp 1\text{ \ .}  \label{1.28b}
\end{eqnarray}
then the above expressions with $\sigma (\xi _{\perp },\xi _{\perp })/$ $%
g(\xi _{\perp },\xi _{\perp })$ could be written in the forms: 
\begin{equation}
v_z=[\frac 1{n-1}\cdot \theta \mp \sigma (n_{\perp },n_{\perp })]\cdot \xi
_{\perp }\text{ ,}  \label{1.28c}
\end{equation}
\begin{equation}
\theta :=0\,:\,\ \ \ \ \ v_z=\mp \sigma (n_{\perp },n_{\perp })\cdot \xi
_{\perp }\text{ \ \ ,}  \label{1.28d}
\end{equation}

The centrifugal (centripetal) relative velocity $v_z$ could be also written
in the form 
\begin{eqnarray}
v_z &=&\frac{g(_{rel}v,\xi _{\perp )}}{g(\xi _{\perp },\xi _{\perp })}\cdot
\xi _{\perp }=\mp \frac{g(_{rel}v,\xi _{\perp )}}{l_{\xi _{\perp }}^2}\cdot
\xi _{\perp }=  \nonumber \\
&=&\mp g(_{rel}v,\frac{\xi _{\perp }}{l_{\xi _{\perp }}})\cdot \frac{\xi
_{\perp }}{l_{\xi _{\perp }}}=\mp g(_{rel}v,n_{\perp })\cdot n_{\perp }\text{
.}  \label{1.28e}
\end{eqnarray}

On the other side, on the basis of the relations 
\begin{eqnarray*}
_{rel}v &=&\overline{g}[d(\xi _{\perp })]=\overline{g}[\sigma (\xi _{\perp
})]+\overline{g}[\omega (\xi _{\perp })]+\frac 1{n-1}\cdot \theta \cdot \xi
_{\perp }\text{ ,} \\
\overline{g}[h_u(\xi )] &=&\xi _{\perp }\text{ \thinspace \thinspace
\thinspace ,\thinspace \thinspace \thinspace \thinspace \thinspace
\thinspace \thinspace }\overline{g}[h_u(\xi _{\perp })]=\overline{g}[h_u(%
\overline{g}[h_u(\xi )])]=\overline{g}[h_u(\xi )]=\xi _{\perp }\text{ ,} \\
g(\overline{g}[\sigma (\xi _{\perp })],n_{\perp }) &=&\sigma (n_{\perp },\xi
_{\perp })=\sigma (n_{\perp },l_{\xi _{\perp }}\cdot n_{\perp })=l_{\xi
_{\perp }}\cdot \sigma (n_{\perp },n_{\perp })\text{ ,} \\
g(\overline{g}[\omega (\xi _{\perp })],n_{\perp }) &=&\omega (n_{\perp },\xi
_{\perp })=\omega (n_{\perp },l_{\xi _{\perp }}\cdot n_{\perp })=l_{\xi
_{\perp }}\cdot \omega (n_{\perp },n_{\perp })=0\text{ \thinspace \thinspace
,} \\
g(\xi _{\perp },n_{\perp }) &=&l_{\xi _{\perp }}\cdot g(n_{\perp },n_{\perp
})=\mp l_{\xi _{\perp }}\text{ ,}
\end{eqnarray*}
it follows for $g(_{rel}v,n_{\perp })$%
\begin{equation}
g(_{rel}v,n_{\perp })=[\sigma (n_{\perp },n_{\perp })\mp \frac 1{n-1}\cdot
\theta ]\cdot l_{\xi _{\perp }}\text{ \thinspace \thinspace ,}  \label{1.28f}
\end{equation}
and for $v_z$%
\begin{eqnarray}
v_z &=&\mp g(_{rel}v,n_{\perp })\cdot n_{\perp }=[\mp \sigma (n_{\perp
},n_{\perp })+\frac 1{n-1}\cdot \theta ]\cdot l_{\xi _{\perp }}\cdot
n_{\perp }\text{ \thinspace \thinspace ,}  \nonumber \\
v_z &=&[\frac 1{n-1}\cdot \theta \mp \sigma (n_{\perp },n_{\perp })]\cdot
l_{\xi _{\perp }}\cdot n_{\perp }\text{ \thinspace \thinspace .}
\label{1.28g}
\end{eqnarray}

Since $l_{\xi _{\perp }}>0$, we have three different cases:

(a) 
\begin{eqnarray*}
v_z &>&0:\mp \sigma (n_{\perp },n_{\perp })+\frac 1{n-1}\cdot \theta >0: \\
\theta &>&\pm (n-1)\cdot \sigma (n_{\perp },n_{\perp })\text{ ,\thinspace
\thinspace \thinspace \thinspace \thinspace \thinspace \thinspace \thinspace
\thinspace \thinspace \thinspace }n-1>0\text{ .}
\end{eqnarray*}

(b) 
\begin{eqnarray*}
v_z &<&0:\mp \sigma (n_{\perp },n_{\perp })+\frac 1{n-1}\cdot \theta <0: \\
\theta &<&\pm (n-1)\cdot \sigma (n_{\perp },n_{\perp })\text{ ,\thinspace
\thinspace \thinspace \thinspace \thinspace \thinspace \thinspace \thinspace
\thinspace \thinspace \thinspace }n-1>0\text{ .}
\end{eqnarray*}

(c) 
\begin{eqnarray*}
v_z &=&0:\mp \sigma (n_{\perp },n_{\perp })+\frac 1{n-1}\cdot \theta =0: \\
\theta &=&\pm (n-1)\cdot \sigma (n_{\perp },n_{\perp })\text{ ,\thinspace
\thinspace \thinspace \thinspace \thinspace \thinspace \thinspace \thinspace
\thinspace \thinspace \thinspace }n-1>0\text{ \thinspace \thinspace
\thinspace \thinspace \thinspace .}
\end{eqnarray*}

\textit{Special case}: $\sigma :=0$ (shear-free relative velocity) 
\[
v_z=\frac 1{n-1}\cdot \theta \cdot l_{\xi _{\perp }}\cdot n_{\perp }\text{
\thinspace \thinspace .} 
\]

For $v_z>0:\theta >0$ we have an expansion, for $v_z<0:\theta <0$ we have a
contraction, and for $v_z=0:\theta =0$ we have a stationary case.

\textit{Special case:} $\theta :=0$ (expansion-free relative velocity) 
\[
v_z=\mp \sigma (n_{\perp },n_{\perp })\cdot l_{\xi _{\perp }}\cdot n_{\perp }%
\text{ \thinspace \thinspace .} 
\]

For $v_z>0:\mp \sigma (n_{\perp },n_{\perp })>0$ we have an expansion, for $%
v_z<0:\mp \sigma (n_{\perp },n_{\perp })<0$ we have a contraction, and for $%
v_z=0:\sigma (n_{\perp },n_{\perp })=0$ we have a stationary case.

In an analogous way we can find the explicit form of $v_z^2$ as 
\begin{eqnarray}
v_z^2 &=&\{\mp [\sigma (n_{\perp },n_{\perp })]^2\mp \frac 1{(n-1)^2}\cdot
\theta ^2+\frac 2{n-1}\cdot \theta \cdot \sigma (n_{\perp },n_{\perp
})\}\cdot l_{\xi _{\perp }}^2=  \label{1.28h} \\
&=&\mp [\sigma (n_{\perp },n_{\perp })\mp \frac 1{n-1}\cdot \theta ]^2\cdot
l_{\xi _{\perp }}^2\text{ .}  \label{1.28i}
\end{eqnarray}

The expression for $v_z^2$ could be now written in the form 
\begin{equation}
v_z^2=\mp H^2\cdot l_{\xi _{\perp }}^2=\mp l_{v_z}^2\text{\thinspace
\thinspace \thinspace \thinspace \thinspace \thinspace \thinspace ,}
\label{1.28j}
\end{equation}
where 
\begin{equation}
H^2=[\sigma (n_{\perp },n_{\perp })\mp \frac 1{n-1}\cdot \theta ]^2\text{ .}
\label{1.28k}
\end{equation}

Then 
\begin{eqnarray}
l_{v_z}^2 &=&H^2\cdot l_{\xi _{\perp }}^2\text{ \thinspace \thinspace
\thinspace \thinspace ,\thinspace \thinspace \thinspace \thinspace
\thinspace \thinspace \thinspace \thinspace \thinspace \thinspace \thinspace
\thinspace }l_{v_z}=\,\mid H\mid \cdot l_{\xi _{\perp }}\text{\thinspace
\thinspace \thinspace \thinspace \thinspace ,\thinspace \thinspace
\thinspace \thinspace \thinspace \thinspace }l_{v_z}>0\text{ \thinspace
\thinspace .}  \label{1.28l} \\
H &=&\mp [\sigma (n_{\perp },n_{\perp })\mp \frac 1{n-1}\cdot \theta ]=
\label{1.28m} \\
&=&\frac 1{n-1}\cdot \theta \mp \sigma (n_{\perp },n_{\perp })\,\,\,\,\,\,%
\text{.}  \label{1.28n}
\end{eqnarray}
and 
\begin{equation}
v_z=\pm \,l_{v_z}\cdot n_{\perp }=\,\pm \,\mid H\mid \cdot l_{\xi _{\perp
}}\cdot n_{\perp }=H\cdot l_{\xi _{\perp }}\cdot n_{\perp }=\,H\cdot \xi
_{\perp }\text{ \thinspace \thinspace \thinspace .}  \label{1.28o}
\end{equation}

It follows from the last expression that the centrifugal (centripetal)
relative velocity $v_z$ is collinear to the vector field $\xi _{\perp }$.
Its absolute value $l_{v_z}=\,\mid H\mid \cdot l_{\xi _{\perp }}$ and the
expression for the centrifugal (centripetal) relative velocity $v_z=H\cdot
l_{\xi _{\perp }}\cdot n_{\perp }=\,H\cdot \xi _{\perp }$ represent
generalizations of the \textit{Hubble law }\cite{Misner}. The function $H$
could be called \textit{Hubble function }(some authors call it \textit{%
Hubble coefficient} \cite{Misner}). Since $H=H(x^k(\tau ))=H(\tau )$, for a
given proper time $\tau =\tau _0$ the function $H$ has at the time $\tau _0$
the value $H(\tau _0)=H(\tau =\tau _0)=\,$const. The Hubble function $H$ is
usually called \textit{Hubble constant}.

\textit{Remark}. The Hubble coefficient $H$ has dimension $sec^{-1}$. The
function $H^{-1}$ with dimension $sec$ is usually denoted in astrophysics as 
\textit{Hubble's time} \cite{Misner}.

The Hubble function $H$ could also be represented in the forms \cite%
{Manoff-7a} 
\begin{eqnarray}
H &=&\frac 1{n-1}\cdot \theta \mp \sigma (n_{\perp },n_{\perp })=\,\, 
\nonumber \\
&=&\frac 12[(n_{\perp })(h_u)(\nabla _u\overline{g}-\pounds _u\overline{g}%
)(h_u)(n_{\perp })]\text{ \thinspace \thinspace \thinspace \thinspace .}
\label{1.28p}
\end{eqnarray}

In the Einstein theory of gravitation (ETG) the Hubble coefficient is
considered under the condition that the centrifugal relative velocity is
generated by a shear-free relative velocity in a cosmological model of the
type of Robertson-Walker \cite{Misner}, i.e. 
\begin{equation}
v_z=H\cdot l_{\xi _{\perp }}\cdot n_{\perp }=H\cdot \xi _{\perp }\text{
\thinspace \thinspace \thinspace \thinspace \thinspace \thinspace
with\thinspace \thinspace \thinspace \thinspace \thinspace }H=\frac
1{n-1}\cdot \theta \text{ \thinspace \thinspace \thinspace \thinspace
\thinspace \thinspace \thinspace \thinspace \thinspace \thinspace and
\thinspace \thinspace \thinspace \thinspace \thinspace }\sigma =0\text{
\thinspace \thinspace \thinspace \thinspace .}  \label{1.28q}
\end{equation}

\textit{Special case:} $U_n$ - or $V_n$-space: $\nabla _u\overline{g}=0$, $%
\pounds _u\overline{g}:=0$ (the vector field $u$ is a Killing vector field
in the corresponding space) 
\[
H=0\text{ \thinspace .} 
\]

Therefore, in a (pseudo) Riemannian space with or without torsion ($U_n$- or 
$V_n$- space) the Hubble function $H$ is equal to zero if the velocity
vector field $u$ ($n=4$) of an observer is a Killing vector field fulfilling
the Killing equation $\pounds _u\overline{g}=0$. This means that the
condition $\pounds _u\overline{g}=0$ is a sufficient condition for the
Hubble function to be equal to zero.

\textit{Special case:} $(\overline{L}_n,g)$-space with $\nabla _u\overline{g}%
-\pounds _u\overline{g}:=0$ \cite{Manoff-15}. For this case, the last
expression appears as a sufficient condition for the vanishing of the Hubble
function $H$, i.e. $H=0$ if $\nabla _u\overline{g}=\pounds _u\overline{g}$.

\textit{Special case}: $(\overline{L}_n,g)$-spaces with vanishing
centrifugal (centripetal) velocity: $v_z:=0$. 
\begin{eqnarray*}
v_z &:&=0:H=0:\frac 1{n-1}\cdot \theta \mp \sigma (n_{\perp },n_{\perp })=0%
\text{ \thinspace \thinspace \thinspace ,} \\
\theta &=&\pm (n-1)\cdot \sigma (n_{\perp },n_{\perp })\text{ \thinspace
\thinspace .}
\end{eqnarray*}

\textit{Remark}. The Hubble function $H$ is introduced in the above
considerations only on a purely kinematic basis related to the notions of
relative velocity and centrifugal (centripetal) relative velocity. Its
dynamic interpretation in a theory of gravitation depends on the structure
of the theory and the relations between the field equations and the Hubble
function.

\subsection{Generalized Hubble's law and centrifugal (centripetal) relative
velocity}

\subsubsection{Null (isotropic) vector fields in spaces with affine
connections and metrics}

The null (isotropic) vector fields are used in all cases where a radiation
process and its corresponding radiation has to be described in relativistic
mechanics and electrodynamics as well as in Einstein's theory of gravitation 
\cite{Stephani}. The null (isotropic) vector fields are related to the wave
vector of the classical electrodynamics and used for an invariant
description of wave propagation in different types of spaces or space-times.

\textit{Definition}. A null (isotropic) vector filed is a contravariant
vector field $k:\neq \mathbf{0\in }T(M)$ with zero length $l_k=\,\mid
g(k,k)\mid ^{1/2}\,=0$.

\textit{Remark}. Since 
\[
k^2=g(k,k)=0=\pm l_k^2\text{ } 
\]
we have $l_k=\,\mid k\mid \,=\,\mid g(k,k)\mid ^{1/2}=0$.

As a contravariant vector field the null vector field $k$ could be
represented by its projections along and orthogonal to a given non-isotropic
(non-null) vector field $u$%
\[
k=\frac 1e\cdot g(u,k)\cdot u+\overline{g}[h_u(k)]\text{ \thinspace
\thinspace \thinspace ,} 
\]
where 
\[
e=g(u,u)=\pm l_u^2:\neq 0\text{ \thinspace \thinspace \thinspace .} 
\]

The invariant (scalar product of $u$ and $k$) $g(u,k):=\omega $ is
interpreted as the frequency $\omega $ of the radiation related to the null
vector $k$ and detected by an observer with velocity $u$%
\begin{equation}
g(u,k)=\omega :=2\cdot \pi \cdot \nu \text{ \thinspace \thinspace \thinspace
\thinspace .}  \label{5.1}
\end{equation}

The contravariant non-isotropic (non-null) vector field $u$ and its
corresponding vector field $\xi _{\perp }$, orthogonal to $u$ [$g(u,\xi
_{\perp })=0$], could be written in the forms 
\begin{eqnarray}
u &=&l_u\cdot n_{\shortparallel }\text{ \thinspace \thinspace \thinspace
\thinspace \thinspace ,\thinspace \thinspace \thinspace \thinspace
\thinspace \thinspace \thinspace \thinspace \thinspace }l_u>0\text{
\thinspace \thinspace ,\thinspace \thinspace \thinspace }  \label{5.2} \\
\text{\thinspace \thinspace \thinspace \thinspace \thinspace \thinspace }%
g(u,u) &=&l_u^2\cdot g(n_{\shortparallel },n_{\shortparallel })=\pm l_u^2%
\text{ ,}  \label{5.2a} \\
g(n_{\shortparallel },n_{\shortparallel }) &=&\pm 1\text{ ,}  \label{5.3} \\
\xi _{\perp } &=&l_{\xi _{\perp }}\cdot n_{\perp }\text{ \thinspace
\thinspace \thinspace ,\thinspace \thinspace \thinspace \thinspace
\thinspace \thinspace \thinspace \thinspace }l_{\xi _{\perp }}>0\text{
\thinspace \thinspace \thinspace ,\thinspace \thinspace \thinspace
\thinspace \thinspace \thinspace }  \label{5.4} \\
g(\xi _{\perp },\xi _{\perp }) &=&l_{\xi _{\perp }}^2\cdot g(n_{\perp
},n_{\perp })=\mp l_{\xi _{\perp }}^2\text{ ,}  \label{5.4a} \\
g(n_{\perp },n_{\perp }) &=&\mp 1\text{ \thinspace .}  \label{5.5}
\end{eqnarray}

The vector fields $n_{\shortparallel }$ and $n_{\perp }$ are normalized unit
vector fields orthogonal to each other, i.e. 
\begin{equation}
g(n_{\shortparallel },n_{\perp })=0\text{ \thinspace \thinspace .}
\label{5.6}
\end{equation}

Proof. From $g(u,\xi _{\perp })=0=g(l_u\cdot n_{\shortparallel },l_{\xi
_{\perp }}\cdot n_{\perp })=l_u\cdot l_{\xi _{\perp }}\cdot
g(n_{\shortparallel },n_{\perp })$ and $l_u\neq 0$, $l_{\xi _{\perp }}\neq 0$%
, it follows that $g(n_{\shortparallel },n_{\perp })=0$.

By the use of the unit vector fields $n_{\shortparallel }$ and $n_{\perp }$
the null vector field $k$ could be written in the form 
\begin{equation}
k=k_{\shortparallel }+k_{\perp }\text{ \thinspace \thinspace \thinspace
,\thinspace \thinspace \thinspace \thinspace \thinspace \thinspace
\thinspace \thinspace \thinspace \thinspace \thinspace }g(k_{\shortparallel
},k_{\perp })=0\text{ \thinspace \thinspace \thinspace \thinspace \thinspace
,}  \label{5.7}
\end{equation}
where 
\begin{equation}
k_{\shortparallel }=\frac \omega e\cdot u=\pm \frac \omega {l_u^2}\cdot
l_u\cdot n_{\shortparallel }=\pm \frac \omega {l_u}\cdot n_{\shortparallel }%
\text{ \thinspace \thinspace \thinspace \thinspace \thinspace .}  \label{5.8}
\end{equation}
\begin{eqnarray}
k_{\perp } &=&\overline{g}[h_u(k)]=k\mp \frac \omega {l_u}\cdot
n_{\shortparallel }=\mp l_{k_{\perp }}\cdot n_{\perp }\text{ \thinspace ,
\thinspace \thinspace \thinspace \thinspace \thinspace \thinspace \thinspace
\thinspace \thinspace \thinspace }  \label{5.9} \\
g(k_{\perp },k_{\perp }) &=&\mp l_{k_{\perp }}^2=l_{k_{\perp }}^2\cdot
g(n_{\perp },n_{\perp })\text{ \thinspace \thinspace ,}  \label{5.10}
\end{eqnarray}
\begin{eqnarray}
g(k,k) &=&g(k_{\perp },k_{\perp })+g(k_{\shortparallel },k_{\shortparallel
})=  \nonumber \\
&=&g(k_{\perp },k_{\perp })+g(\pm \frac \omega {l_u}\cdot n_{\shortparallel
},\pm \frac \omega {l_u}\cdot n_{\shortparallel })=  \nonumber \\
&=&\mp l_{k_{\perp }}^2+\frac{\omega ^2}{l_u^2}\cdot g(n_{\shortparallel
},n_{\shortparallel })=\mp l_{k_{\perp }}^2\pm \frac{\omega ^2}{l_u^2}=0%
\text{ \thinspace \thinspace \thinspace ,}  \label{5.11}
\end{eqnarray}
\begin{equation}
\mp l_{k_{\perp }}^2=\mp \frac{\omega ^2}{l_u^2}\text{ \thinspace \thinspace
\thinspace \thinspace ,\thinspace \thinspace \thinspace \thinspace
\thinspace \thinspace \thinspace \thinspace \thinspace \thinspace }%
l_{k_{\perp }}^2=\text{\thinspace \thinspace }\frac{\omega ^2}{l_u^2}\text{
\thinspace \thinspace \thinspace \thinspace \thinspace ,\thinspace
\thinspace \thinspace \thinspace \thinspace \thinspace \thinspace \thinspace
\thinspace \thinspace }l_{k_{\perp }}=\frac \omega {l_u}\text{ \thinspace
\thinspace \thinspace \thinspace ,\thinspace }  \label{5.12}
\end{equation}
\begin{equation}
k_{\perp }=\mp \frac \omega {l_u}\cdot n_{\perp }\,\,\,\,\,\,\,\text{,}
\label{5.13}
\end{equation}
\begin{equation}
k=\frac \omega {l_u}\cdot (\pm n_{\shortparallel }\mp n_{\perp })\text{
\thinspace \thinspace \thinspace ,\thinspace \thinspace \thinspace
\thinspace \thinspace \thinspace \thinspace \thinspace \thinspace }g(k,k)=0%
\text{ \thinspace \thinspace \thinspace .}  \label{5.14}
\end{equation}

\subsubsection{Centrifugal (centripetal) relative velocity and null vector
fields}

The frequency $\omega $ of the radiation related to the null vector field $k$
and detected by an observer with velocity vector field $u$ has been
determined as the projection of the non-isotropic (non-null) velocity vector
field $u$ at the vector field $k$ (or vice versa): $\omega =g(u,k)$. In an
analogous way, the projection of the centrifugal (centripetal) relative
velocity $v_z$ at the part of $k$, orthogonal to $u$, (or vice versa) could
be interpreted as the change $_{rel}\omega :=g(v_z,k_{\perp })$ of the
frequency $\overline{\omega }$ of the emitter under the motion of the
observer described by the centrifugal (centripetal) relative velocity $v_z$.
By the use of the relations 
\begin{eqnarray*}
g(u,k) &=&\omega \text{ \thinspace \thinspace \thinspace \thinspace
\thinspace ,\thinspace \thinspace \thinspace \thinspace \thinspace
\thinspace \thinspace \thinspace \thinspace \thinspace \thinspace \thinspace
\thinspace }g(n_{\perp },k_{\perp })=\frac \omega {l_u}\text{\thinspace
\thinspace \thinspace \thinspace \thinspace \thinspace \thinspace ,} \\
v_z &=&[\frac 1{n-1}\cdot \theta \mp \sigma (n_{\perp },n_{\perp })]\cdot
l_{\xi _{\perp }}\cdot n_{\perp }\text{ \thinspace \thinspace ,}
\end{eqnarray*}
the projection $g(v_z,k_{\perp })$ of $v_z$ at $k_{\perp }$ could be found
in the form 
\begin{eqnarray}
_{rel}\omega &=&g(v_z,k_{\perp })=[\frac 1{n-1}\cdot \theta \mp \sigma
(n_{\perp },n_{\perp })]\cdot l_{\xi _{\perp }}\cdot g(n_{\perp },k_{\perp
})=  \label{5.15} \\
&=&\frac \omega {l_u}\cdot [\frac 1{n-1}\cdot \theta \mp \sigma (n_{\perp
},n_{\perp })]\cdot l_{\xi _{\perp }}\text{ \thinspace \thinspace \thinspace
\thinspace \thinspace .}  \label{5.16}
\end{eqnarray}
If we introduce the abbreviation for the Hubble function 
\begin{equation}
H=\frac 1{n-1}\cdot \theta \mp \sigma (n_{\perp },n_{\perp })\,
\label{5.16a}
\end{equation}
then the expression for $_{rel}\omega $ could be written in the form 
\begin{equation}
_{rel}\omega =H\cdot \frac{l_{\xi _{\perp }}}{l_u}\cdot \omega \text{
\thinspace \thinspace \thinspace \thinspace \thinspace \thinspace \thinspace
\thinspace .}  \label{5.17}
\end{equation}

As mentioned above, the frequency $\omega $ is the frequency detected by the
observer with the velocity vector field $u$. The frequency of the emitter $%
\overline{\omega }$ could be find by the use of the relation between $\omega 
$ and $\overline{\omega }$ on the basis of the expression of the change of $%
\omega $ under the centrifugal (centripetal) motion [described by the
centrifugal (centripetal) velocity $v_z$] as 
\begin{equation}
\overline{\omega }=\omega +\,_{rel}\omega \text{\thinspace \thinspace
\thinspace \thinspace \thinspace \thinspace \thinspace ,\thinspace
\thinspace \thinspace \thinspace \thinspace \thinspace \thinspace \thinspace
\thinspace \thinspace \thinspace \thinspace \thinspace }\omega =\overline{%
\omega }-\,_{rel}\omega \text{\thinspace \thinspace \thinspace \thinspace
\thinspace \thinspace \thinspace .}  \label{5.18}
\end{equation}

Therefore, the radiation with frequency $\overline{\omega }$ by the emitter
could be expressed as 
\begin{equation}
\overline{\omega }=\omega +\,_{rel}\omega =\omega +H\cdot \frac{l_{\xi
_{\perp }}}{l_u}\cdot \omega =(1+H\cdot \frac{l_{\xi _{\perp }}}{l_u})\cdot
\omega  \label{15.19}
\end{equation}
and it will be detected by the observer as the frequency $\omega $. The
relative difference between both the frequencies (emitted $\overline{\omega }
$ and detected $\omega $) 
\[
\frac{\overline{\omega }-\omega }\omega :=\frac{\bigtriangleup \omega }%
\omega  
\]
appears in the form 
\begin{equation}
\frac{\bigtriangleup \omega }\omega =\frac{\overline{\omega }-\omega }\omega
=H\cdot \frac{l_{\xi _{\perp }}}{l_u}\,\,\,\,\,\,\text{.}  \label{5.20}
\end{equation}

If we introduce the abbreviation 
\begin{equation}
z:=H\cdot \frac{l_{\xi _{\perp }}}{l_u}\text{ }  \label{5.21}
\end{equation}
we obtain the relation between the emitted frequency $\overline{\omega }$
and the frequency detected by the observer in the form 
\begin{equation}
\frac{\overline{\omega }-\omega }\omega =z\,\,\,\,\,\,\,\,\text{,\thinspace
\thinspace \thinspace \thinspace \thinspace \thinspace \thinspace \thinspace
\thinspace \thinspace \thinspace \thinspace }\overline{\omega }=(1+z)\cdot
\omega \text{\thinspace \thinspace \thinspace \thinspace \thinspace .}
\label{5.22}
\end{equation}

The quantity $z$ could be denoted as \textit{observed shift frequency
parameter}.

If $z=0$ then $H=0$ and there will be no difference between the emitted and
detected frequencies: $\overline{\omega }=\omega $, i.e. for $z=0$ $%
\overline{\omega }=\omega $. This will be the case if 
\begin{equation}
\theta =\pm (n-1)\cdot \sigma (n_{\perp },n_{\perp })\,\,\,\,\,\text{.}
\label{5.23}
\end{equation}

If $z>0$ the observed shift frequency parameter is called \textit{red shift}%
. If $z<0$ the observed shift frequency parameter is called \textit{blue
shift}. If $\overline{\omega }$ and $\omega $ are known the observed shift
frequency parameter $z$ could be found. If $w$ and $z$ are given then the
corresponding $\overline{\omega }$ could be estimated.

On the other side, from the explicit form of $z$%
\begin{equation}
z=H\cdot \frac{l_{\xi _{\perp }}}{l_u}=[\frac 1{n-1}\cdot \theta \mp \sigma
(n_{\perp },n_{\perp })]\cdot \frac{l_{\xi _{\perp }}}{l_u}  \label{5.24}
\end{equation}
we could find the relation between the observed shift frequency parameter $z$
and the kinematic characteristics of the relative velocity such as the
expansion and shear velocities.

\textit{Special case}: $(\overline{L}_n,g)$-spaces with shear-free relative
velocity: $\sigma :=0$. 
\begin{equation}
z=\frac 1{n-1}\cdot \theta \cdot \frac{l_{\xi _{\perp }}}{l_u}\text{
\thinspace \thinspace \thinspace \thinspace \thinspace \thinspace
,\thinspace \thinspace \thinspace \thinspace \thinspace \thinspace
\thinspace \thinspace \thinspace \thinspace \thinspace \thinspace \thinspace
\thinspace }H=\frac 1{n-1}\cdot \theta \,\,\,\,\,\,\,\,\text{.}  \label{5.25}
\end{equation}

\textit{Special case}: $(\overline{L}_n,g)$-spaces with expansion-free
relative velocity: $\theta :=0$. 
\begin{equation}
z=\mp \sigma (n_{\perp },n_{\perp })\cdot \frac{l_{\xi _{\perp }}}{l_u}%
\,\,\,\,\,\,\,\,\,\text{,\thinspace \thinspace \thinspace \thinspace
\thinspace \thinspace \thinspace \thinspace \thinspace \thinspace \thinspace 
}H=\mp \sigma (n_{\perp },n_{\perp })\,\,\,\,\,\,\,\text{.}  \label{5.26}
\end{equation}

On the grounds of the observed shift frequency parameter $z$ the distance
(the length $l_{\xi _{\perp }}$ of $\xi _{\perp }$) between the observer
[with the world line $x^i(\tau )$ and velocity vector field $u=\frac d{d\tau
}$] and the observed object (at a distance $l_{\xi _{\perp }}$ from the
observer) emitted radiation with null vector field $k$ could be found as 
\begin{equation}
l_{\xi _{\perp }}=z\cdot \frac{l_u}H=\frac{\overline{\omega }-\omega }\omega
\cdot \frac{l_u}H\,\,\,\,\,\,\text{.}  \label{5.27}
\end{equation}

On the other side, if $z$, $H$, and $l_{\xi _{\perp }}$ are known the
absolute value $l_u$ of the velocity vector $u$ could be found as 
\begin{equation}
l_u=\frac Hz\cdot l_{\xi _{\perp }}=\frac{H\cdot \omega }{\overline{\omega }%
-\omega }\cdot l_{\xi _{\perp }}\text{\thinspace \thinspace \thinspace
\thinspace \thinspace \thinspace \thinspace \thinspace .}  \label{5.28}
\end{equation}

\textit{Remark}. In the Einstein theory of gravitation (ETG) the absolute
value of $u$ is usually normalized to $1$ or $c$, i.e. $l_u=1,c$. Then the
last expression could be used for experimental check up of the velocity $c$
of light in vacuum if $z$, $H$, and $l_{\xi _{\perp }}$ are known 
\begin{equation}
c=\frac Hz\cdot l_{\xi _{\perp }}=\frac{H\cdot \omega }{\overline{\omega }%
-\omega }\cdot l_{\xi _{\perp }}\,\,\,\,\,\,\text{.}  \label{5.29}
\end{equation}

Since 
\[
z=[\frac 1{n-1}\cdot \theta \mp \sigma (n_{\perp },n_{\perp })]\cdot \frac{%
l_{\xi _{\perp }}}{l_u}=\frac{\overline{\omega }-\omega }\omega  
\]
it follows that 
\begin{equation}
l_u=\frac \omega {\overline{\omega }-\omega }\cdot [\frac 1{n-1}\cdot \theta
\mp \sigma (n_{\perp },n_{\perp })]\cdot l_{\xi _{\perp }}\,\,\,\,\,\,\text{.%
}  \label{5.30}
\end{equation}

Analogous expression we can find for the length $l_{\xi _{\perp }}$ of the
vector field $\xi _{\perp }$%
\begin{equation}
l_{\xi _{\perp }}=(n-1)\cdot (\frac{\overline{\omega }}\omega -1)\cdot \frac{%
l_u}{\theta \mp (n-1)\cdot \sigma (n_{\perp },n_{\perp })}\,\,\,\,\,\,\,%
\text{.}  \label{5.31}
\end{equation}

\textit{Special case}: $(\overline{L}_n,g)$-space with shear-free relative
velocity: $\sigma :=0$. 
\begin{equation}
l_u=\frac \omega {\overline{\omega }-\omega }\cdot \frac 1{n-1}\cdot \theta
\cdot l_{\xi _{\perp }}\,\,\,\,\,\,\text{,}  \label{5.32}
\end{equation}
\begin{equation}
l_{\xi _{\perp }}=(n-1)\cdot (\frac{\overline{\omega }}\omega -1)\cdot \frac{%
l_u}\theta \,\,\,\,\,\,\,\,\text{.}  \label{5.33}
\end{equation}

\textit{Special case}: $(\overline{L}_n,g)$-space with expansion-free
relative velocity: $\theta :=0$. 
\begin{equation}
l_u=\mp \frac \omega {\overline{\omega }-\omega }\cdot \sigma (n_{\perp
},n_{\perp })\cdot l_{\xi _{\perp }}\,\,\,\,\,\,\,\,\text{,}  \label{5.34}
\end{equation}
\begin{equation}
l_{\xi _{\perp }}=\mp (\frac{\overline{\omega }}\omega -1)\cdot \frac{l_u}{%
\sigma (n_{\perp },n_{\perp })}\,\,\,\,\,\,\,\,\,\,\text{.}  \label{5.35}
\end{equation}

By the use of the relation between the Hubble function $H$ and the observed
shift parameter $z$ we can express the centrifugal (centripetal) velocity by
means of the frequencies $\overline{\omega }$ and $\omega $. From 
\[
v_z=H\cdot l_{\xi _{\perp }}\cdot n_{\perp }\text{\thinspace \thinspace
\thinspace \thinspace \thinspace \thinspace \thinspace ,\thinspace
\thinspace \thinspace \thinspace \thinspace \thinspace \thinspace \thinspace
\thinspace \thinspace \thinspace \thinspace \thinspace \thinspace \thinspace
\thinspace }H=z\cdot \frac{l_u}{l_{\xi _{\perp }}}=(\frac{\overline{\omega }}%
\omega -1)\cdot \frac{l_u}{l_{\xi _{\perp }}}\text{ \thinspace \thinspace
\thinspace ,} 
\]
it follows that 
\begin{equation}
v_z=z\cdot l_u\cdot n_{\perp }=(\frac{\overline{\omega }}\omega -1)\cdot
l_u\cdot n_{\perp }\,\,\,\,\,\,\text{.}  \label{5.36}
\end{equation}

Then 
\[
g(v_z,v_z)=\mp (\frac{\overline{\omega }}\omega -1)^2\cdot l_u^2=\mp
l_{v_z}^2\text{ \thinspace \thinspace \thinspace \thinspace \thinspace ,} 
\]
\begin{eqnarray}
l_{v_z} &=&\pm (\frac{\overline{\omega }}\omega -1)\cdot l_u\text{
\thinspace \thinspace \thinspace \thinspace \thinspace \thinspace \thinspace
,\thinspace \thinspace \thinspace \thinspace \thinspace \thinspace
\thinspace \thinspace \thinspace \thinspace \thinspace }\pm l_{v_z}=(\frac{%
\overline{\omega }}\omega -1)\cdot l_u\,\,\,\,\,\text{,}  \label{5.37} \\
l_{v_z} &>&0\text{ \thinspace \thinspace \thinspace ,\thinspace \thinspace
\thinspace \thinspace \thinspace \thinspace \thinspace \thinspace \thinspace 
}l_u>0\text{,\thinspace \thinspace \thinspace \thinspace \thinspace
\thinspace \thinspace \thinspace \thinspace }l_{\xi _{\perp
}}>0\,\,\,\,\,\,\,\text{,}  \label{3.37a}
\end{eqnarray}
where (since $l_{v_z}>0$) 
\begin{eqnarray}
\overline{\omega } &>&\omega :l_{v_z}=(\frac{\overline{\omega }}\omega
-1)\cdot l_u\text{\thinspace \thinspace \thinspace \thinspace \thinspace
\thinspace \thinspace \thinspace \thinspace \thinspace ,}  \label{5.38} \\
\overline{\omega } &<&\omega :l_{v_z}=(1-\frac{\overline{\omega }}\omega
)\cdot l_u\,\,\,\,\,\,\,\,\,\text{.}  \label{5.39}
\end{eqnarray}

On the other side, we can express the relation between $\overline{\omega }$
and $\omega $ by means of the last relations: 
\begin{equation}
\frac{\overline{\omega }}\omega =1\pm \frac{l_{v_z}}{l_u}\text{\thinspace
\thinspace \thinspace \thinspace \thinspace \thinspace \thinspace \thinspace
\thinspace \thinspace ,\thinspace \thinspace \thinspace \thinspace
\thinspace \thinspace \thinspace \thinspace \thinspace \thinspace \thinspace 
}\overline{\omega }=(1\pm \frac{l_{v_z}}{l_u})\cdot \omega \,\,\,\,\,\,\,\,%
\text{,}  \label{5.40}
\end{equation}
\begin{eqnarray}
\overline{\omega } &>&\omega :\text{\thinspace \thinspace }\overline{\omega }%
=(1+\frac{l_{v_z}}{l_u})\cdot \omega \,\,\,\,\,\,\,\text{,}  \label{5.41a} \\
\overline{\omega } &<&\omega :\overline{\omega }=(1-\frac{l_{v_z}}{l_u}%
)\cdot \omega \,\,\,\,\,\,\,\,\text{.}  \label{5.41b}
\end{eqnarray}

If we express the frequencies as $\overline{\omega }=2\cdot \pi \cdot 
\overline{\nu }$ and $\omega =2\cdot \pi \cdot \nu $ we obtain 
\begin{eqnarray}
\overline{\nu } &>&\nu :\text{\thinspace \thinspace }\overline{\nu }=(1+%
\frac{l_{v_z}}{l_u})\cdot \nu \,\,\,\,\,\,\,\text{,}  \label{5.42a} \\
\overline{\nu } &<&\omega :\overline{\nu }=(1-\frac{l_{v_z}}{l_u})\cdot \nu
\,\,\,\,\,\,\,\,\text{.}  \label{5.42b}
\end{eqnarray}

The last relations represent a generalization of the \textit{Doppler effect
in }$(\overline{L}_n,g)$\textit{-spaces}.

It should be stressed that the generalized Doppler effect is a result of
pure kinematic considerations of the properties of a null (isotropic) vector
field by means of the kinematic characteristics of the relative velocity in
spaces with affine connections and metrics. The Hubble function $H$, the
observed shift frequency parameter $z$ are kinematic characteristics related
to the centrifugal (centripetal) relative velocity. For different classic
field theories they could have different relations to the dynamic variables
of the corresponding theory.

\subsection{Coriolis' velocity}

The vector field 
\begin{equation}
v_{c}=\overline{g}[h_{\xi _{\perp }}(_{rel}v)]=g^{ij}\cdot (h_{\xi _{\perp
}})_{\overline{j}\overline{k}}\cdot ~_{rel}v^{k}\cdot \partial _{i}\text{ }
\label{1.28}
\end{equation}
is called Coriolis' velocity.

\paragraph{Properties of the Coriolis' velocity}

(a) The Coriolis velocity is orthogonal to the vector field $u$, interpreted
as velocity of a mass element (particle), i.e. 
\begin{equation}
g(u,v_{c})=0\text{ .}  \label{1.29}
\end{equation}

Proof: From the definition of the Coriolis velocity, it follows 
\begin{eqnarray}
g(u,v_{c}) &=&g(u,\overline{g}[h_{\xi _{\perp }}(_{rel}v)]=g_{\overline{i}%
\overline{j}}\cdot u^{i}\cdot g^{jk}\cdot (h_{\xi _{\perp }})_{\overline{k}%
\overline{l}}\cdot ~_{rel}v^{l}=  \nonumber \\
&=&g_{\overline{i}\overline{j}}\cdot g^{jk}\cdot u^{i}\cdot (h_{\xi _{\perp
}})_{\overline{k}\overline{l}}\cdot ~_{rel}v^{l}=g_{i}^{k}\cdot u^{i}\cdot
(h_{\xi _{\perp }})_{\overline{k}\overline{l}}\cdot ~_{rel}v^{l}=
\label{1.30} \\
&=&(h_{\xi _{\perp }})_{\overline{k}\overline{l}}\cdot u^{k}\cdot
~_{rel}v^{l}=(u)(h_{\xi _{\perp }})(_{rel}v)\text{ .}  \nonumber
\end{eqnarray}

Since 
\begin{eqnarray}
(u)(h_{\xi _{\perp }}) &=&(h_{\xi _{\perp }})_{\overline{k}\overline{l}%
}\cdot u^k\cdot dx^l=(u)(g)-\frac 1{g(\xi _{\perp },\xi _{\perp })}\cdot
(u)[g(\xi _{\perp })]\cdot g(\xi _{\perp })=  \nonumber \\
&=&g(u)-\frac 1{g(\xi _{\perp },\xi _{\perp })}\cdot g(u,\xi _{\perp })\cdot
g(\xi _{\perp })=g(u)\text{ \ \ , \ }  \label{1.31} \\
\text{\ \ }g(u,\xi _{\perp }) &=&0\text{ \ ,\ }  \nonumber
\end{eqnarray}
then $(u)(h_{\xi _{\perp }})(_{rel}v)=[g(u)](_{rel}v)=g(u,_{rel}v)=0$.
Because of $g(u,_{rel}v)=0$, it follows that $g(u,v_c)=0$.

(b) The Coriolis velocity $v_{c}$ is orthogonal to the centrifugal
(centripetal) velocity $v_{z}$%
\begin{equation}
g(v_{c},v_{z})=0\text{ .}  \label{1.32}
\end{equation}

(c) The length of the vector $v_{c}$ could be found by the use of the
relations: 
\begin{eqnarray}
v_{c} &=&\overline{g}[h_{\xi _{\perp }}(_{rel}v)]~\ \ \ \ \text{,}  \nonumber
\\
h_{\xi _{\perp }}(_{rel}v) &=&g(_{rel}v)-\frac{1}{g(\xi _{\perp },\xi
_{\perp })}\cdot \lbrack g(\xi _{\perp })](_{rel}v)\cdot g(\xi _{\perp })= 
\nonumber \\
&=&g(_{rel}v)-\frac{g(\xi _{\perp },_{rel}v)}{g(\xi _{\perp },\xi _{\perp })}%
\cdot g(\xi _{\perp })\text{ \ ,}  \label{1.33} \\
\lbrack g(\xi _{\perp })](_{rel}v) &=&g(\xi _{\perp },_{rel}v)\text{ ,} 
\nonumber
\end{eqnarray}

\begin{equation}
v_{c}=\overline{g}[h_{\xi _{\perp }}(_{rel}v)]=~_{rel}v-\frac{g(\xi _{\perp
},_{rel}v)}{g(\xi _{\perp },\xi _{\perp })}\cdot \xi _{\perp }\text{ \ \ \ ,}
\label{1.34}
\end{equation}
\begin{eqnarray*}
g(v_{c},v_{c}) &=&v_{c}^{2}=g(_{rel}v-\frac{g(\xi _{\perp },_{rel}v)}{g(\xi
_{\perp },\xi _{\perp })}\cdot \xi _{\perp },_{rel}v-\frac{g(\xi _{\perp
},_{rel}v)}{g(\xi _{\perp },\xi _{\perp })}\cdot \xi _{\perp })= \\
&=&g(_{rel}v,_{rel}v)-\frac{g(\xi _{\perp },_{rel}v)}{g(\xi _{\perp },\xi
_{\perp })}\cdot g(\xi _{\perp },_{rel}v)-\frac{g(\xi _{\perp },_{rel}v)}{%
g(\xi _{\perp },\xi _{\perp })}\cdot g(_{rel}v,\xi _{\perp })+ \\
&&+\frac{[g(\xi _{\perp },_{rel}v)]^{2}}{g(\xi _{\perp },\xi _{\perp })}%
\text{ ,}
\end{eqnarray*}
\begin{equation}
g(v_{c},v_{c})=v_{c}^{2}=g(_{rel}v,_{rel}v)-\frac{[g(\xi _{\perp
},_{rel}v)]^{2}}{g(\xi _{\perp },\xi _{\perp })}\text{ \ .}  \label{1.35}
\end{equation}

On the other side, because of $(u)(h_{u})=h_{u}(u)=0$, and $_{rel}v=%
\overline{g}[h_{u}(\nabla _{u}\xi _{\perp })]$,~~it follows that 
\[
g(u,_{rel}v)=h_{u}(u,\nabla _{u}\xi _{\perp })=(u)(h_{u})(\nabla _{u}\xi
_{\perp })\text{ \ ,} 
\]

\begin{equation}
g(u,_{rel}v)=(u)(h_{u})(\nabla _{u}\xi _{\perp })=0\text{ .}  \label{1.36}
\end{equation}

Since 
\begin{eqnarray}
g(_{rel}v,_{rel}v) &=&~_{rel}v^{2}=g(\overline{g}[d(\xi _{\perp })],%
\overline{g}[d(\xi _{\perp })])\text{=}  \nonumber \\
&=&\overline{g}(d(\xi _{\perp }),d(\xi _{\perp }))\text{ \ \ \ , \ \ \ \ }%
\pounds _{u}\xi _{\perp }=0\text{ \ ,}  \label{1.37}
\end{eqnarray}
\begin{eqnarray*}
\overline{g}(d(\xi _{\perp }),d(\xi _{\perp })) &=&\overline{g}(\sigma (\xi
_{\perp })+\omega (\xi _{\perp })+\frac{1}{n-1}\cdot \theta \cdot g(\xi
_{\perp }), \\
&&\sigma (\xi _{\perp })+\omega (\xi _{\perp })+\frac{1}{n-1}\cdot \theta
\cdot g(\xi _{\perp }))
\end{eqnarray*}
we obtain

\begin{eqnarray}
g(_{rel}v,_{rel}v) &=&~_{rel}v^{2}=\overline{g}(d(\xi _{\perp }),d(\xi
_{\perp }))=  \nonumber \\
&=&\overline{g}(\sigma (\xi _{\perp }),\sigma (\xi _{\perp }))+\overline{g}%
(\omega (\xi _{\perp }),\omega (\xi _{\perp }))+  \nonumber \\
&&+\frac{1}{(n-1)^{2}}\cdot \theta ^{2}\cdot g(\xi _{\perp },\xi _{\perp })+
\nonumber \\
&&+2\cdot \overline{g}(\sigma (\xi _{\perp }),\omega (\xi _{\perp }))+\frac{2%
}{n-1}\cdot \theta \cdot \sigma (\xi _{\perp },\xi _{\perp })\text{ .}
\label{1.38}
\end{eqnarray}

\textit{Special case:} $\sigma :=0$, $\theta :=0$ (shear-free and
expansion-free velocity). 
\begin{equation}
_{rel}v^{2}=\overline{g}(\omega (\xi _{\perp }),\omega (\xi _{\perp }))\text{
.}  \label{1.39}
\end{equation}

\textit{Special case:} $\omega :=0$ (rotation-free velocity). 
\begin{equation}
_{rel}v^{2}=\overline{g}(\sigma (\xi _{\perp }),\sigma (\xi _{\perp }))+%
\frac{1}{(n-1)^{2}}\cdot \theta ^{2}\cdot g(\xi _{\perp },\xi _{\perp })+%
\frac{2}{n-1}\cdot \theta \cdot \sigma (\xi _{\perp },\xi _{\perp })\text{ .}
\label{1.40}
\end{equation}

\textit{Special case:} $\theta :=0$, $\omega :=0$ (expansion-free and
rotation-free velocity). 
\begin{equation}
_{rel}v^{2}=\overline{g}(\sigma (\xi _{\perp }),\sigma (\xi _{\perp }))\text{
\ .}  \label{1.41}
\end{equation}

\textit{Special case:} $\sigma :=0$, $\omega :=0$ (shear-free and
rotation-free velocity). 
\begin{equation}
_{rel}v^{2}=\frac{1}{(n-1)^{2}}\cdot \theta ^{2}\cdot g(\xi _{\perp },\xi
_{\perp })\text{ \ .}  \label{1.42}
\end{equation}

\textit{Special case:} $\theta :=0$ (expansion-free velocity). 
\begin{equation}
_{rel}v^{2}=\overline{g}(\sigma (\xi _{\perp }),\sigma (\xi _{\perp }))+%
\overline{g}(\omega (\xi _{\perp }),\omega (\xi _{\perp }))+2\cdot \overline{%
g}(\sigma (\xi _{\perp }),\omega (\xi _{\perp }))\text{ .}  \label{1.43}
\end{equation}

The square $v_{c}^{2}$ of $v_{c}$ could be found on the basis of the
relation $v_{c}^{2}=~_{rel}v^{2}-v_{z}^{2}$%
\begin{eqnarray}
v_{c}^{2} &=&~_{rel}v^{2}-v_{z}^{2}=\overline{g}(\sigma (\xi _{\perp
}),\sigma (\xi _{\perp }))-\frac{[\sigma (\xi _{\perp },\xi _{\perp })]^{2}}{%
g(\xi _{\perp },\xi _{\perp })}+  \nonumber \\
&&+\overline{g}(\omega (\xi _{\perp }),\omega (\xi _{\perp }))+2\cdot 
\overline{g}(\sigma (\xi _{\perp }),\omega (\xi _{\perp }))\text{ .}
\label{1.44}
\end{eqnarray}

Therefore, the length of $v_{c}$ does not depend on the expansion velocity
invariant $\theta $.

\textit{Special case:} $\sigma :=0$ (shear-free velocity). 
\begin{equation}
v_{c}^{2}=\overline{g}(\omega (\xi _{\perp }),\omega (\xi _{\perp }))\text{ .%
}  \label{1.45}
\end{equation}

\textit{Special case:} $\omega :=0$ (rotation-free velocity). 
\begin{equation}
v_{c}^{2}=\overline{g}(\sigma (\xi _{\perp }),\sigma (\xi _{\perp }))-\frac{%
[\sigma (\xi _{\perp },\xi _{\perp })]^{2}}{g(\xi _{\perp },\xi _{\perp })}%
\text{ .}  \label{1.46}
\end{equation}

Therefore, even if the rotation velocity tensor $\omega $ is equal to zero $%
(\omega =0)$, the Coriolis (relative) velocity $v_{c}$ is not equal to zero
if $\sigma \neq 0$.

(d) The scalar product between $v_{c}$ and $_{rel}v$ could be found in its
explicit form by the use of the relations: 
\begin{equation}
g(v_{c},_{rel}v)=h_{\xi _{\perp }}(_{rel}v,_{rel}v)=~_{rel}v^{2}-\frac{%
[g(\xi _{\perp },_{rel}v)]^{2}}{g(\xi _{\perp },\xi _{\perp })}%
=~_{rel}v^{2}-v_{z}^{2}=v_{c}^{2}\text{ .}  \label{1.47}
\end{equation}

(e) The explicit form of $v_{c}$ could be found by the use of the relations 
\[
v_{c}=\overline{g}[h_{\xi _{\perp }}(_{rel}v)]=~_{rel}v-\frac{g(\xi _{\perp
},_{rel}v)}{g(\xi _{\perp },\xi _{\perp })}\cdot \xi _{\perp }=~_{rel}v-v_{z}%
\text{ ,} 
\]
\begin{equation}
v_{c}=\overline{g}[\sigma (\xi _{\perp })]-\frac{\sigma (\xi _{\perp },\xi
_{\perp })}{g(\xi _{\perp },\xi _{\perp })}\cdot \xi _{\perp }+\overline{g}%
[\omega (\xi _{\perp })]\text{ .}  \label{1.48}
\end{equation}

Therefore, the Coriolis velocity does not depend on the expansion velocity
invariant $\theta $.

\textit{Special case:} $\sigma :=0$ (shear-free velocity). 
\begin{equation}
v_{c}=\overline{g}[\omega (\xi _{\perp })]\text{ .}  \label{1.49}
\end{equation}

\textit{Special case:} $\omega :=0$ (rotation-free velocity). 
\begin{equation}
v_{c}=\overline{g}[\sigma (\xi _{\perp })]-\frac{\sigma (\xi _{\perp },\xi
_{\perp })}{g(\xi _{\perp },\xi _{\perp })}\cdot \xi _{\perp }\text{ .}
\label{1.50}
\end{equation}

(f) The Coriolis velocity $v_{c}$ is orthogonal to the deviation vector $\xi
_{\perp }$, i.e. $g(v_{c},\xi _{\perp })=0$.

Proof: From $g(v_{c},\xi _{\perp })=g(\overline{g}[h_{\xi _{\perp
}}(_{rel}v)],\xi _{\perp })=h_{\xi _{\perp }}(\xi _{\perp },_{rel}v)=(\xi
_{\perp })(h_{\xi _{\perp }})(_{rel}v)$ and $(\xi _{\perp })(h_{\xi _{\perp
}})=0$, it follows that 
\begin{equation}
g(v_{c},\xi _{\perp })=0.  \label{1.51}
\end{equation}

\section{Centrifugal (centripetal) and Coriolis' accelerations}

In analogous way as in the case of centrifugal (centripetal) and Coriolis'
velocities, the corresponding accelerations could be defined by the use of
the projections of the relative acceleration $_{rel}a=\overline{g}%
[h_{u}(\nabla _{u}\nabla _{u}\xi _{\perp })]$ along or orthogonal to the
vector field $\xi _{\perp }$%
\begin{equation}
_{rel}a=\frac{g(\xi _{\perp },~_{rel}a)}{g(\xi _{\perp },\xi _{\perp })}%
\cdot \xi _{\perp }+\overline{g}[h_{\xi _{\perp }}(_{rel}a)]=a_{z}+a_{c}%
\text{ \ ,}  \label{2.1}
\end{equation}
where 
\begin{equation}
a_{z}=\frac{g(\xi _{\perp },~_{rel}a)}{g(\xi _{\perp },\xi _{\perp })}\cdot
\xi _{\perp }\text{ \ ,}  \label{2.2}
\end{equation}
\begin{equation}
a_{c}=\overline{g}[h_{\xi _{\perp }}(_{rel}a)]\text{ \ .}  \label{2.3}
\end{equation}

If 
\begin{equation}
\frac{g(\xi _{\perp },~_{rel}a)}{g(\xi _{\perp },\xi _{\perp })}>0
\label{2.4}
\end{equation}
the vector field $a_{z}$ is called (relative) \textit{centrifugal
acceleration.} If 
\begin{equation}
\frac{g(\xi _{\perp },~_{rel}a)}{g(\xi _{\perp },\xi _{\perp })}<0
\label{2.5}
\end{equation}
the vector field $a_{z}$ is called (relative)\textit{\ centripetal
acceleration.}

The vector field $a_{c}$ is called (relative) \textit{Coriolis' acceleration.%
}

\subsection{Centrifugal (centripetal) acceleration}

The relative acceleration $_{rel}a$ is orthogonal to the vector field $u.$

Proof: From $g(u,_{rel}a)=g(u,\overline{g}[h_{u}(\nabla _{u}\nabla _{u}\xi
_{\perp })])=(u)(h_{u})(\nabla _{u}\nabla _{u}\xi _{\perp })$ and $%
(u)(h_{u})=h_{u}(u)=0$, it follows that 
\begin{equation}
g(u,_{rel}a)=0\text{ .}  \label{2.6}
\end{equation}

(a) The centrifugal (centripetal) acceleration $a_{z}$ is orthogonal to the
vector field $u$, i.e. 
\begin{equation}
g(u,a_{z})=0.  \label{2.6a}
\end{equation}

Proof: From 
\begin{eqnarray}
g(u,a_{z}) &=&g(u,\frac{g(\xi _{\perp },~_{rel}a)}{g(\xi _{\perp },\xi
_{\perp })}\cdot \xi _{\perp })=  \nonumber \\
&=&\frac{g(\xi _{\perp },~_{rel}a)}{g(\xi _{\perp },\xi _{\perp })}\cdot
g(u,\xi _{\perp })\text{ , ~}  \label{2.7} \\
\text{\ \ \ \ }g(u,\xi _{\perp }) &=&0\text{ ,}  \nonumber
\end{eqnarray}
it follows that $g(u,a_{z})=0$.

(b) The centrifugal (centripetal) acceleration $a_{z}$ is orthogonal to the
Coriolis acceleration $a_{c}$, i.e. 
\begin{equation}
g(a_{z},a_{c})=0\text{ .}  \label{2.8}
\end{equation}

Proof: From 
\[
g(\frac{g(\xi _{\perp },~_{rel}a)}{g(\xi _{\perp },\xi _{\perp })}\cdot \xi
_{\perp },\overline{g}[h_{\xi _{\perp }}(_{rel}a)])=\frac{g(\xi _{\perp
},~_{rel}a)}{g(\xi _{\perp },\xi _{\perp })}\cdot g(\xi _{\perp },\overline{g%
}[h_{\xi _{\perp }}(_{rel}a)])\text{ ,} 
\]
\begin{eqnarray}
g(\xi _{\perp },\overline{g}[h_{\xi _{\perp }}(_{rel}a)]) &=&(\xi _{\perp
})(h_{\xi _{\perp }})(_{rel}a)\text{ ,}  \label{2.9} \\
(\xi _{\perp })(h_{\xi _{\perp }}) &=&(h_{\xi _{\perp }})(\xi _{\perp })=0%
\text{ ,}  \label{2.10}
\end{eqnarray}
it follows that $g(a_{z},a_{c})=0$.

(c) The length of the vector $a_{z}$ could be found on the basis of the
relations 
\begin{equation}
a_{z}^{2}=g(a_{z},a_{z})=\frac{[g(\xi _{\perp },~_{rel}a)]^{2}}{g(\xi
_{\perp },\xi _{\perp })}\text{ \ \ \ .}  \label{2.11}
\end{equation}

Therefore, in general, the square $a_{z}^{2}$ of the length of the
centrifugal (centripetal) acceleration $a_{z}$ is reverse proportional to $%
\xi _{\perp }^{2}=g(\xi _{\perp },\xi _{\perp })$.

\textit{Special case:} $M_n=E_n$, $n=3$ (3-dimensional Euclidean space). 
\begin{eqnarray}
\xi _{\perp } &:&=\overrightarrow{r}\text{ , \ \ \ }g(\xi _{\perp },\xi
_{\perp })=r^2\text{ \ , \ \ \ \ }_{rel}a=~_{rel}\overrightarrow{a}\text{\ },
\nonumber \\
a_z^2 &=&\frac{[g(\overrightarrow{r},~_{rel}\overrightarrow{a})]^2}{r^2}%
\text{ , \ \ \ \ \ }l_{a_z}=\frac{g(\overrightarrow{r},\,_{rel}%
\overrightarrow{a})}r=g(n_{\perp },\,_{rel}\overrightarrow{a})\text{ ,}
\label{2.12} \\
n_{\perp } &:&=\frac{\overrightarrow{r}}r\text{ , \ \ \ \ \ \ }g(n_{\perp
},n_{\perp })=\frac{g(\overrightarrow{r},\overrightarrow{r})}{r^2}=n_{\perp
}^2=1\text{ \ \ .}  \label{2.12a}
\end{eqnarray}

The length $l_{a_z}=\mid g(a_z,a_z)\mid ^{1/2}$ of the centrifugal
(centripetal) acceleration $a_z$ is equal to the projection of the relative
acceleration $_{rel}\overrightarrow{a}$ at the unit vector field $n_{\perp }$
along the vector field $\xi _{\perp }$. If \ $l_{a_z}=g(n_{\perp },\,_{rel}%
\overrightarrow{a})=0$, i.e. if the relative acceleration $_{rel}%
\overrightarrow{a}$ is orthogonal to the radius vector $\overrightarrow{r}$ [%
$_{rel}\overrightarrow{a}\perp \overrightarrow{r}$], then $a_z=0$.

If the relative acceleration $_{rel}a$ is equal to zero then the centrifugal
(centripetal) acceleration $a_{z}$ is also equal to zero.

(d) The scalar product $g(\xi _{\perp },_{rel}a)$ could be found in its
explicit form by the use of the explicit form of $_{rel}a$ \cite{Manoff-5} 
\begin{equation}
_{rel}a=\overline{g}[A(\xi _{\perp })]=\overline{g}[_sD(\xi _{\perp })]+%
\overline{g}[W(\xi _{\perp })]+\frac 1{n-1}\cdot U\cdot \xi _{\perp }\text{ .%
}  \label{2.13}
\end{equation}

Then 
\begin{eqnarray}
g(\xi _{\perp },~_{rel}a) &=&~_{s}D(\xi _{\perp },\xi _{\perp })+\frac{1}{n-1%
}\cdot U\cdot g(\xi _{\perp },\xi _{\perp })\text{ ,}  \label{2.14} \\
a_{z}^{2} &=&\frac{[g(\xi _{\perp },~_{rel}a)]^{2}}{g(\xi _{\perp },\xi
_{\perp })}=  \nonumber \\
&=&\frac{1}{g(\xi _{\perp },\xi _{\perp })}\cdot \lbrack _{s}D(\xi _{\perp
},\xi _{\perp })+\frac{1}{n-1}\cdot U\cdot g(\xi _{\perp },\xi _{\perp
})]^{2}=  \nonumber \\
&=&\frac{[_{s}D(\xi _{\perp },\xi _{\perp })]^{2}}{g(\xi _{\perp },\xi
_{\perp })}+\frac{1}{(n-1)^{2}}\cdot U^{2}\cdot g(\xi _{\perp },\xi _{\perp
})+  \nonumber \\
&&+\frac{2}{n-1}\cdot U\cdot ~_{s}D(\xi _{\perp },\xi _{\perp })\text{ .}
\label{2.15}
\end{eqnarray}

\textit{Special case:} $_{s}D:=0$ (shear-free acceleration). 
\begin{equation}
a_{z}^{2}=\frac{1}{(n-1)^{2}}\cdot U^{2}\cdot g(\xi _{\perp },\xi _{\perp })%
\text{ \ .}  \label{2.16}
\end{equation}

\textit{Special case:} $U:=0$ (expansion-free acceleration). 
\begin{equation}
a_{z}^{2}=\frac{[_{s}D(\xi _{\perp },\xi _{\perp })]^{2}}{g(\xi _{\perp
},\xi _{\perp })}\text{ \ .}  \label{2.17}
\end{equation}

(e) The explicit form of $a_z$ could be found in the form 
\begin{eqnarray}
a_z &=&\frac{g(\xi _{\perp },~_{rel}a)}{g(\xi _{\perp },\xi _{\perp })}\cdot
\xi _{\perp }=\frac{_sD(\xi _{\perp },\xi _{\perp })}{g(\xi _{\perp },\xi
_{\perp })}\cdot \xi _{\perp }+\frac 1{n-1}\cdot U\cdot \xi _{\perp }= 
\nonumber \\
&=&[\frac 1{n-1}\cdot U+\frac{_sD(\xi _{\perp },\xi _{\perp })}{g(\xi
_{\perp },\xi _{\perp })}]\cdot \xi _{\perp }=  \label{2.18} \\
&=&[\frac 1{n-1}\cdot U\pm \,_sD(n_{\perp },n_{\perp })]\cdot \xi _{\perp }%
\text{ \ \ \ .}  \label{2.18a}
\end{eqnarray}

If 
\begin{equation}
\frac{1}{n-1}\cdot U+\frac{_{s}D(\xi _{\perp },\xi _{\perp })}{g(\xi _{\perp
},\xi _{\perp })}>0\text{ }  \label{2.19}
\end{equation}

$a_{z}$ is a centrifugal (relative) acceleration. If 
\begin{equation}
\frac{1}{n-1}\cdot U+\frac{_{s}D(\xi _{\perp },\xi _{\perp })}{g(\xi _{\perp
},\xi _{\perp })}<0  \label{2.20}
\end{equation}

$a_{z}$ is a centripetal (relative) acceleration.

In a theory of gravitation the centripetal (relative) acceleration could be
interpreted as gravitational acceleration.

\textit{Special case:} $_{s}D:=0$ (shear-free relative acceleration). 
\begin{equation}
a_{z}=\frac{1}{n-1}\cdot U\text{ }\cdot \xi _{\perp }\text{ \ \ \ \ .}
\label{2.22}
\end{equation}

If the expansion acceleration invariant $U>0$ the acceleration $a_{z}$ is a
centrifugal (or expansion) acceleration. If $U<0$ the acceleration $a_{z}$
is a centripetal (or contraction) acceleration. Therefore, in the case of a
shear-free relative acceleration the centrifugal or the centripetal
acceleration is proportional to the expansion acceleration invariant $U$.

\subsection{Coriolis' acceleration}

The vector field $a_{c}$, defined as 
\[
a_{c}=\overline{g}[h_{\xi _{\perp }}(_{rel}a)]\text{ ,} 
\]
is called Coriolis' (relative) acceleration. On the basis of its definition,
the Coriolis acceleration has well defined properties.

(a) The Coriolis acceleration is orthogonal to the vector field $u$, i.e. 
\begin{equation}
g(u,a_{c})=0.  \label{2.23}
\end{equation}

Proof: From 
\[
g(u,a_{c})=g(u,\overline{g}[h_{\xi _{\perp }}(_{rel}a)])=(u)(h_{\xi _{\perp
}})(_{rel}a)=h_{\xi _{\perp }}(u,_{rel}a) 
\]
and 
\begin{equation}
(u)(h_{\xi _{\perp }})=(h_{\xi _{\perp }})(u)=g(u)\text{ \ , \ \ \ \ }%
g(u,_{rel}a)=0\text{ \ ,}  \label{2.23a}
\end{equation}
it follows that 
\[
g(u,a_{c})=[g(u)](_{rel}a)=g(u,_{rel}a)=0. 
\]

(b) The Coriolis acceleration $a_{c}$ is orthogonal to the centrifugal
(centripetal) acceleration $a_{z}$, i.e. 
\begin{equation}
g(a_{c},a_{z})=0\text{ \ .}  \label{2.24}
\end{equation}

(c) The Coriolis acceleration $a_{c}$ is orthogonal to the deviation vector $%
\xi _{\perp }$, i.e. 
\begin{equation}
g(\xi _{\perp },a_{c})=0\text{ \ \ .}  \label{2.25}
\end{equation}

(d) The length $\sqrt{\mid a_{c}^{2}\mid }=\sqrt{\mid g(a_{c},a_{c})\mid }$
of $a_{c}$ could be found by the use of the relations 
\begin{eqnarray}
a_{c} &=&\overline{g}[h_{\xi _{\perp }}(_{rel}a)]\text{ ,}  \nonumber \\
\lbrack g(\xi _{\perp })](_{rel}a) &=&g(\xi _{\perp },_{rel}a)\text{ \ ,}
\label{2.26} \\
h_{\xi _{\perp }}(_{rel}a) &=&g(_{rel}a)-\frac{g(\xi _{\perp },_{rel}a)}{%
g(\xi _{\perp },\xi _{\perp })}\cdot g(\xi _{\perp })\text{ \ ,}
\label{2.27}
\end{eqnarray}

\begin{eqnarray}
a_{c} &=&\overline{g}[h_{\xi _{\perp }}(_{rel}a)]=~_{rel}a-a_{z}=
\label{2.28} \\
&=&~_{rel}a-\frac{g(\xi _{\perp },_{rel}a)}{g(\xi _{\perp },\xi _{\perp })}%
\cdot \xi _{\perp }\text{ \ \ \ . }  \label{2.29}
\end{eqnarray}

For $a_{c}^{2}$ we obtain 
\begin{eqnarray}
a_{c}^{2} &=&g(a_{c},a_{c})=g(_{rel}a-a_{z},~_{rel}a-a_{z})=  \nonumber \\
&=&g(_{rel}a,_{rel}a)+g(a_{z},a_{z})-2\cdot g(a_{z},_{rel}a)\text{ \ .}
\label{2.30}
\end{eqnarray}

Since 
\begin{equation}
g(a_{z},a_{z})=g(a_{z},_{rel}a)=\frac{[g(\xi _{\perp }),_{rel}a)]^{2}}{g(\xi
_{\perp },\xi _{\perp })\text{ }}\text{ \ \ ,}  \label{2.31}
\end{equation}
\begin{equation}
a_{c}^{2}=g(_{rel}a,_{rel}a)-g(a_{z},a_{z})=~_{rel}a^{2}-a_{z}^{2}\text{ \ \
,}  \label{2.32}
\end{equation}
\begin{eqnarray*}
g(_{rel}a,_{rel}a) &=&~_{rel}a^{2}=g(\overline{g}[A(\xi _{\perp })],%
\overline{g}[A(\xi _{\perp })])= \\
&=&\overline{g}(A(\xi _{\perp }),A(\xi _{\perp }))\text{ ,} \\
A &=&~_{s}D+W+\frac{1}{n-1}\cdot U\cdot h_{u}\text{ \ \ ,} \\
A(\xi _{\perp }) &=&~_{s}D(\xi _{\perp })+W(\xi _{\perp })+\frac{1}{n-1}%
\cdot U\cdot g(\xi _{\perp })\text{ \ ,} \\
h_{u}(\xi _{\perp }) &=&g(\xi _{\perp })\text{ \ ,}
\end{eqnarray*}

\begin{eqnarray}
\overline{g}(_{s}D(\xi _{\perp }),h_{u}(\xi _{\perp })) &=&~_{s}D(\xi
_{\perp },\xi _{\perp })\text{ \ ,}  \label{2.33} \\
\overline{g}(W(\xi _{\perp }),h_{u}(\xi _{\perp })) &=&~W(\xi _{\perp },\xi
_{\perp })=0\text{ \ \ ,}  \label{2.34} \\
\overline{g}(h_{u}(\xi _{\perp }),h_{u}(\xi _{\perp })) &=&~h_{u}(\xi
_{\perp },\xi _{\perp })=g\text{ }(\xi _{\perp },\xi _{\perp })\text{ , }
\label{2.35}
\end{eqnarray}
it follows for $_{rel}a^{2}$%
\begin{eqnarray}
_{rel}a^{2} &=&g(_{rel}a,_{rel}a)=\overline{g}(_{s}D(\xi _{\perp
}),_{s}D(\xi _{\perp }))+\overline{g}(W(\xi _{\perp }),W(\xi _{\perp }))+ 
\nonumber \\
&&+2\cdot \overline{g}(_{s}D(\xi _{\perp }),W(\xi _{\perp }))+  \nonumber \\
&&+\frac{2}{n-1}\cdot U\cdot ~_{s}D(\xi _{\perp },\xi _{\perp })+\frac{1}{%
(n-1)^{2}}\cdot U^{2}\cdot g\text{ }(\xi _{\perp },\xi _{\perp })\text{ .}
\label{2.36}
\end{eqnarray}

On the other side, 
\[
a_{z}^{2}=\frac{[_{s}D(\xi _{\perp },\xi _{\perp })]^{2}}{g(\xi _{\perp
},\xi _{\perp })}+\frac{1}{(n-1)^{2}}\cdot U^{2}\cdot g(\xi _{\perp },\xi
_{\perp })+\frac{2}{n-1}\cdot U\cdot ~_{s}D(\xi _{\perp },\xi _{\perp })%
\text{ .} 
\]

Therefore, 
\begin{eqnarray}
a_{c}^{2} &=&~_{rel}a^{2}-a_{z}^{2}=  \nonumber \\
&=&\overline{g}(_{s}D(\xi _{\perp }),_{s}D(\xi _{\perp }))-\frac{[_{s}D(\xi
_{\perp },\xi _{\perp })]^{2}}{g(\xi _{\perp },\xi _{\perp })}+  \nonumber \\
&&+\overline{g}(W(\xi _{\perp }),W(\xi _{\perp }))+2\cdot \overline{g}%
(_{s}D(\xi _{\perp }),W(\xi _{\perp }))\text{ \ .}  \label{2.37}
\end{eqnarray}

\textit{Special case:} $_{s}D:=0$ (shear-free acceleration). 
\begin{equation}
a_{c}^{2}=\overline{g}(W(\xi _{\perp }),W(\xi _{\perp }))\text{ \ \ .}
\label{2.38}
\end{equation}

\textit{Special case:} $W:=0$ (rotation-free acceleration). 
\begin{equation}
a_{c}^{2}=\overline{g}(_{s}D(\xi _{\perp }),_{s}D(\xi _{\perp }))-\frac{%
[_{s}D(\xi _{\perp },\xi _{\perp })]^{2}}{g(\xi _{\perp },\xi _{\perp })}%
\text{ \ \ .}  \label{2.39}
\end{equation}

The explicit form of $a_c$ could be found by the use of the relations: 
\begin{eqnarray}
a_c &=&\overline{g}[h_{\xi _{\perp }}(_{rel}a)]=~_{rel}a-\frac{g(\xi _{\perp
},_{rel}a)}{g(\xi _{\perp },\xi _{\perp })}\cdot \xi _{\perp }=~_{rel}a-a_z=
\nonumber \\
&=&\overline{g}[_sD(\xi _{\perp })]-\frac{_sD(\xi _{\perp },\xi _{\perp })}{%
g(\xi _{\perp },\xi _{\perp })}\cdot \xi _{\perp }+\overline{g}[W(\xi
_{\perp })]=  \label{2.40} \\
&=&\overline{g}[_sD(\xi _{\perp })]\mp \,_sD(n_{\perp },n_{\perp })\cdot \xi
_{\perp }+\overline{g}[W(\xi _{\perp })]\text{ .}  \label{2.40a}
\end{eqnarray}

Therefore, the Coriolis acceleration $a_{c}$ does not depend on the
expansion acceleration invariant $U$.

\textit{Special case:} $_{s}D:=0$ (shear-free acceleration). 
\begin{equation}
a_{c}=\overline{g}[W(\xi _{\perp })]\text{ \ \ .}  \label{2.41}
\end{equation}

\textit{Special case:} $W:=0$ (rotation-free acceleration). 
\begin{eqnarray}
a_c &=&\overline{g}[_sD(\xi _{\perp })]-\frac{_sD(\xi _{\perp },\xi _{\perp
})}{g(\xi _{\perp },\xi _{\perp })}\cdot \xi _{\perp }=  \label{2.42} \\
&=&\overline{g}[_sD(\xi _{\perp })]\mp \,_sD(n_{\perp },n_{\perp })\cdot \xi
_{\perp }\text{ \ .}  \label{2.42a}
\end{eqnarray}

The Coriolis acceleration depends on the shear acceleration $_{s}D$ and on
the rotation acceleration $W$. These types of accelerations generate a
Coriolis acceleration between particles or mass elements in a flow.

\section{Centrifugal (centripetal) acceleration as gravitational acceleration%
}

1. The main idea of the Einstein theory of gravitation (ETG) is the
identification of the centripetal acceleration with the gravitational
acceleration. The weak equivalence principle stays that a gravitational
acceleration could be compensated by a centripetal acceleration and vice
versa. From this point of view, it is worth to be investigated the relation
between the centrifugal (centripetal) acceleration and the Einstein theory
of gravitation as well as the possibility for describing the gravitational
interaction as result of the centrifugal (centripetal) acceleration
generated by the motion of mass elements (particles).

The structure of the centrifugal (centripetal) acceleration could be
considered on the basis of its explicit form expressed by means of the
kinematic characteristics of the relative acceleration and the relative
velocity. The centrifugal (centripetal) acceleration is written as 
\[
a_{z}=[\frac{1}{n-1}\cdot U+\frac{_{s}D(\xi _{\perp },\xi _{\perp })}{g(\xi
_{\perp },\xi _{\perp })}]\cdot \xi _{\perp }\text{ .} 
\]

The vector field $\xi _{\perp }$ is directed outside of the trajectory of a
mass element (particle). If the mass element generates a gravitational field
the acceleration $a_{z}$ should be in the direction to the mass element. If
the mass element moves in an external gravitational field caused by an other
gravitational source then the acceleration $a_{z}$ should be directed to the
source. The expansion (contraction) invariant $U$ could be expressed by
means of the kinematic characteristics of the relative acceleration or of
the relative velocity in the forms \cite{Manoff-3}, \cite{Manoff-5} 
\begin{equation}
U=U_{0}+I=~_{F}U_{0}-~_{T}U_{0}+I\text{ \ , \ \ \ \ \ \ \ \ }%
U_{0}=~_{F}U_{0}-~_{T}U_{0}\text{ ,}~  \label{3.1}
\end{equation}
where $_{F}U_{0}$ is the torsion-free and curvature-free expansion
acceleration, $_{T}U_{0}$ is the expansion acceleration induced by the
torsion and $I$ is the expansion acceleration induced by the curvature, $%
U_{0}$ is the curvature-free expansion acceleration 
\begin{eqnarray}
_{F}U_{0} &=&g[b]-\frac{1}{e}\cdot g(u,\nabla _{u}a)\text{ ,}  \label{3.2} \\
_{F}U_{0} &=&a^{k}~_{;k}-\frac{1}{e}\cdot g_{\overline{k}\overline{l}}\cdot
u^{k}\cdot a^{l}~_{;m}\cdot u^{m}\text{ ,}  \label{3.3}
\end{eqnarray}
\begin{eqnarray}
U_{0} &=&g[b]-\overline{g}[_{s}P(\overline{g})\sigma ]-\overline{g}[Q(%
\overline{g})\omega ]-\dot{\theta}_{1}-\frac{1}{n-1}\cdot \theta _{1}\cdot
\theta -  \nonumber \\
&&-\frac{1}{e}\cdot \lbrack g(u,T(a,u))+g(u,\nabla _{u}a)]\text{ ,}
\label{3.4}
\end{eqnarray}
\begin{equation}
I=R_{ij}\cdot u^{i}\cdot u^{j}\text{ \ , \ \ \ \ \ }g[b]=g_{\overline{i}%
\overline{j}}\cdot b^{ij}=g_{\overline{i}\overline{j}}\cdot a^{i}~_{;n}\cdot
g^{nj}=a^{n}~_{;n}\text{ .}  \label{3.5}
\end{equation}

2. In the ETG only the term $I$ is used on the basis of the Einstein
equations. The invariant $I$ represents an invariant generalization of
Newton's gravitational law \cite{Manoff-8}. In a $V_{n}$-space $(n=4)$ of
the ETG, a free moving spinless test particle with $a=0$ will have an
expansion (contraction) acceleration $U=I$ $(U_{0}=0)$ if $R_{ij}\neq 0$ and 
$U=I=0$ if $R_{ij}=0$. At the same time, in a $V_{n}$-space (the bars over
the indices should be omitted)

\begin{eqnarray}
_{s}D &=&~_{s}M\text{ , \ \ \ \ \ }_{s}D_{0}=0\text{ , \ \ \ \ }%
M=h_{u}(K_{s})h_{u}\text{ ,}  \label{3.6} \\
M_{ij} &=&h_{i\overline{k}}\cdot K_{s}^{kl}\cdot h_{\overline{l}j}=\frac{1}{2%
}\cdot h_{i\overline{k}}\cdot (K_{s}^{kl}+K_{s}^{lk})\cdot h_{\overline{l}j}=
\nonumber \\
&=&\frac{1}{2}\cdot h_{i\overline{k}}\cdot (R^{k}~_{mnr}\cdot u^{m}\cdot
u^{n}\cdot g^{rl}+R^{l}~_{mnr}\cdot u^{m}\cdot u^{n}\cdot g^{rk})\cdot h_{%
\overline{l}j}=  \nonumber \\
&=&\frac{1}{2}\cdot u^{m}\cdot u^{n}\cdot (h_{i\overline{k}}\cdot
R^{k}~_{mnr}\cdot g^{rl}\cdot h_{\overline{l}j}+h_{i\overline{k}}\cdot
R^{l}~_{mnr}\cdot g^{rk}\cdot h_{\overline{l}j})=  \nonumber \\
&=&\frac{1}{2}\cdot u^{m}\cdot u^{n}\cdot \lbrack h_{i\overline{k}}\cdot
R^{k}~_{mnr}\cdot g^{rl}\cdot (g_{\overline{l}j}-\frac{1}{e}\cdot u_{%
\overline{l}}\cdot u_{j})+  \nonumber \\
&&+h_{\overline{l}j}\cdot R^{k}~_{mnr}\cdot g^{rk}\cdot (g_{\overline{k}i}-%
\frac{1}{e}\cdot u_{\overline{k}}\cdot u_{i})\text{ \ .}  \label{3.7}
\end{eqnarray}

Since 
\[
h_{i\overline{k}}\cdot R^k~_{mnr}\cdot g^{rl}\cdot (g_{\overline{l}j}-\frac
1e\cdot u_{\overline{l}}\cdot u_j)= 
\]
\begin{eqnarray*}
&=&(g_{i\overline{k}}-\frac 1e\cdot u_i\cdot u_{\overline{k}})\cdot
R^k~_{mnr}\cdot g^{rl}\cdot g_{\overline{l}j}- \\
&&-\frac 1e\cdot (g_{i\overline{k}}-\frac 1e\cdot u_i\cdot u_{\overline{k}%
})\cdot R^k~_{mnr}\cdot g^{rl}\cdot u_{\overline{l}}\cdot u_j\text{ \ ,}
\end{eqnarray*}
\begin{eqnarray*}
R^k~_{mnr}\cdot g^{rl}\cdot u_{\overline{l}} &=&R^k~_{mnr}\cdot u^r\text{ \ ,%
} \\
u_{\overline{k}}\cdot R^k~_{mnr} &=&g_{\overline{k}\overline{s}}\cdot
u^s\cdot R^k~_{mnr}\text{ \ ,}
\end{eqnarray*}
we have

\begin{eqnarray}
h_{i\overline{k}}\cdot R^k~_{mnr}\cdot g^{rl}\cdot h_{\overline{l}j} &=&g_{i%
\overline{k}}\cdot R^k~_{mnr}\cdot g^{rl}\cdot g_{\overline{l}j}-  \nonumber
\\
&&-\frac 1e\cdot u_i\cdot g_{\overline{k}\overline{s}}\cdot u^s\cdot
R^k~_{mnr}\cdot g^{rl}\cdot g_{\overline{l}j}-  \nonumber \\
&&-\frac 1e\cdot g_{i\overline{k}}\cdot R^k~_{mnr}\cdot g^{rl}\cdot g_{%
\overline{l}\overline{s}}\cdot u^s\cdot u_j+  \label{(3.8)} \\
&&+\frac 1{e^2}\cdot u_i\cdot g_{\overline{k}\overline{s}}\cdot u^s\cdot
R^k~_{mnr}\cdot g^{rl}\cdot g_{\overline{l}\overline{q}}\cdot u^q\cdot u_j%
\text{ .}  \nonumber
\end{eqnarray}

\textit{Special case:} $V_{n}$-space: $S:=C$. 
\begin{equation}
M_{ij}=R_{imnj}\cdot u^{m}\cdot u^{n}\text{ \ \ \ \ .}  \label{3.9}
\end{equation}
\begin{eqnarray}
_{s}D_{ij} &=&~_{s}M_{ij}=M_{ij}-\frac{1}{n-1}\cdot I\cdot h_{ij}=  \nonumber
\\
&=&R_{imnj}\cdot u^{m}\cdot u^{n}-\frac{1}{n-1}\cdot R_{mn}\cdot u^{m}\cdot
u^{n}\cdot h_{ij\text{ }}\text{,}  \nonumber \\
_{s}D_{ij} &=&(R_{imnj}-\frac{1}{n-1}\cdot R_{mn}\cdot h_{ij\text{ }})\cdot
u^{m}\cdot u^{n}=~_{s}M_{ij}\text{ ,}  \label{3.10} \\
U &=&I=R_{ij}\cdot u^{i}\cdot u^{j}\text{ .}  \nonumber
\end{eqnarray}

In a $V_{n}$-space the components $a_{z}^{i}$ of the centrifugal
(centripetal) acceleration $a_{z}$ have the form 
\begin{eqnarray}
a_{z}^{i} &=&(\frac{1}{n-1}\cdot U+\frac{_{s}D_{jk}\cdot \xi _{\perp
}^{j}\cdot \xi _{\perp }^{k}}{g_{rs}\cdot \xi _{\perp }^{r}\cdot \xi _{\perp
}^{s}})\cdot \xi _{\perp }^{i}=  \label{3.11} \\
&=&[\frac{1}{n-1}\cdot R_{mn}\cdot u^{m}\cdot u^{n}+  \nonumber \\
&&+\frac{1}{g_{rs}\cdot \xi _{\perp }^{r}\cdot \xi _{\perp }^{s}}\cdot
(R_{jmnk}-\frac{1}{n-1}\cdot R_{mn}\cdot h_{jk})\cdot u^{m}\cdot u^{n}\cdot
\xi _{\perp }^{j}\cdot \xi _{\perp }^{k}]\cdot \xi _{\perp }^{i}\text{ .} 
\nonumber
\end{eqnarray}

3. If the Einstein equations in vacuum without cosmological term $(\lambda
_{0}=0)$ are valid, i.e. if 
\begin{equation}
R_{ij}=0\text{ \ , \ \ }\lambda _{0}=\text{const.}=0\text{ \ , \ \ \ \ }n=4%
\text{ ,}  \label{3.12}
\end{equation}
are fulfilled then 
\begin{eqnarray}
a_{z}^{i} &=&\frac{1}{g_{rs}\cdot \xi _{\perp }^{r}\cdot \xi _{\perp }^{s}}%
\cdot R_{jmnk}\cdot u^{m}\cdot u^{n}\cdot \xi _{\perp }^{j}\cdot \xi _{\perp
}^{k}\cdot \xi _{\perp }^{i}=  \label{3.13} \\
&=&R_{jmnk}\cdot u^{m}\cdot u^{n}\cdot n^{j}\cdot n^{k}\cdot \xi _{\perp
}^{i}=\mathbf{g}\cdot \xi _{\perp }^{i}\text{ ,}  \label{3.14}
\end{eqnarray}
where 
\begin{equation}
n^{j}=\frac{\xi _{\perp }^{j}}{\sqrt{\mid g_{rs}\cdot \xi _{\perp }^{r}\cdot
\xi _{\perp }^{s}\mid }}\text{ ,}  \label{3.15}
\end{equation}
\begin{equation}
\mathbf{g=}R_{jmnk}\cdot u^{m}\cdot u^{n}\cdot n^{j}\cdot n^{k}\text{ \ .}
\label{3.16}
\end{equation}

If $a_{z}$ is a centripetal acceleration interpreted as gravitational
acceleration for a free spinless test particles moving in an external
gravitational field $(R_{ij}=0)$ then the condition $\mathbf{g<~}0$ should
be valid if \ the centripetal acceleration is directed to the particle. If
the centripetal acceleration is directed to the gravitational source (in the
direction $\xi _{\perp }$) then $\mathbf{g>~}0$.

From a more general point of view as that in the ETG, a gravitational theory
could be worked out in a $(\overline{L}_{n},g)$-space where $a_{z}$ could
also be interpreted as gravitational acceleration of mass elements or
particles generating a gravitational field by themselves and caused by their
motions in space-time.

4. If we consider a frame of reference in which a mass element (particle) is
at rest then $u^{i}=g_{4}^{i}\cdot u^{4}$ and $\xi _{\perp
}^{i}=g_{a}^{i}\cdot \xi _{\perp }^{a}:=g_{1}^{i}\cdot \xi _{\perp }^{1}$.
The centrifugal (centripetal) acceleration $a_{z}^{i}$ could be written in
the form 
\begin{equation}
a_{z}^{i}=R_{1441}\cdot u^{4}\cdot u^{4}\cdot n^{1}\cdot n^{1}\cdot \xi
_{\perp }^{i}=R_{1441}\cdot (u^{4})^{2}\cdot (n^{1})^{2}\cdot \xi _{\perp
}^{i}\text{ \ .}  \label{3.17}
\end{equation}

For the Schwarzschild metric 
\begin{eqnarray}
ds^2 &=&\frac{dr^2}{1-\frac{r_g}r}+r^2\cdot (d\theta ^2+\sin ^2\theta \cdot
d\varphi ^2)-(1-\frac{r_g}r)\cdot (dx^4)^2\text{ \ , }  \label{3.18} \\
\text{\ }r_g &=&\frac{2\cdot k\cdot M_0}{c^2}\text{ ,}  \label{3.18a}
\end{eqnarray}
the component $R_{1441}=g_{11}\cdot R^1~_{441}$ of the curvature tensor has
the form \cite{Stephani} 
\begin{equation}
R_{1441}=-g_{11}\cdot (\Gamma _{44,1}^1-\Gamma _{14,4}^1+\Gamma _{11}^1\cdot
\Gamma _{44}^1+\Gamma _{14}^1\cdot \Gamma _{44}^4-\Gamma _{14}^1\cdot \Gamma
_{14}^1-\Gamma _{44}^1\cdot \Gamma _{41}^4)\text{ .}  \label{3.19}
\end{equation}

After introducing in the last expression the explicit form of the metric and
of the Christoffel symbols $\Gamma _{jk}^i$, it follows for $R_{1441}$%
\begin{equation}
R_{1441}=\frac{r_g}{r^3}\text{ \ .}  \label{3.20}
\end{equation}

Then 
\begin{equation}
a_z^i=\frac{r_g}{r^3}\cdot (u^4)^2\cdot (n^1)^2\cdot \xi _{\perp }^i\text{ .}
\label{3.21}
\end{equation}

If the co-ordinate time $t=x^4/c$ is chosen as equal to the proper time $%
\tau $ of the particle, i.e. if $t=\tau $ then 
\begin{eqnarray}
u^4 &=&\frac{dx^4}{d\tau }=c\cdot \frac{d\tau }{d\tau }=c\text{ , \ \ \ \ \ }%
n^1=1\text{ \ \ , \ \ \ \ \ \ }\xi _{\perp }^i=g_1^i\cdot \xi _{\perp }^1%
\text{ \ ,}  \label{3.22} \\
a_z^i &=&\frac{r_g}{r^3}\cdot c^2\cdot g_1^i\cdot \xi _{\perp }^1\text{ , \
\ \ \ \ }\frac{r_g}{r^3}\cdot c^2>0\text{ \ \ .}  \label{3.23}
\end{eqnarray}

For the centrifugal (centripetal) \ acceleration we obtain 
\begin{equation}
a_z^1=\frac{2\cdot k\cdot M_0}{r^3}\cdot \xi _{\perp }^1\text{ \ \ ,}
\label{3.24}
\end{equation}
which is exactly the relative gravitational acceleration between two mass
elements (particles) with co-ordinates $x_{c1}=r$ and $x_{c2}=r+\xi _{\perp
}^1$ \cite{Manoff-9}.

Therefore, the centrifugal (centripetal) acceleration could be used for
working out of a theory of gravitation in a space with affine connections
and metrics as this has been done in the Einstein theory of gravitation.

\section{Conclusions}

In the present paper the notions of centrifugal (centripetal) and Coriolis'
velocities and accelerations are introduced and considered in spaces with
affine connections and metrics as velocities and accelerations of flows of
mass elements (particles) moving in space-time. It is shown that these types
of velocities and accelerations are generated by the relative motions
between the mass elements. The null (isotropic) vector fields are considered
and their relations with the centrifugal (centripetal) velocity are
established. The centrifugal (centripetal) velocity is found to be in
connection with the Hubble law and the generalized Doppler effect in spaces
with affine connections and metrics. The accelerations are closely related
to the kinematic characteristics of the relative velocity and relative
acceleration. The centrifugal (centripetal) acceleration could be
interpreted as gravitational acceleration as it has been done in the
Einstein theory of gravitation. This fact could be used as a basis for
working out of new gravitational theories in spaces with affine connections
and metrics.

\end{document}